\documentclass[superscriptaddress,amsmath,amssymb,aps,prb,twocolumn,floatfix]{revtex4-2}

\usepackage{graphicx}
\usepackage{physics}
\usepackage{bm}
\usepackage[colorlinks, linkcolor=mycol, citecolor=mycol, urlcolor=mycol, breaklinks]{hyperref}
\usepackage{braket}
\usepackage{bbold}
\usepackage{bm}
\usepackage[normalem]{ulem}
\usepackage{tabularx, colortbl}
\usepackage{comment}
\usepackage{appendix}
\usepackage{xcolor}
\usepackage{dsfont}
\usepackage{enumitem}

\definecolor{mycol}{RGB}{10,55,130}

\begin{document}

\title{Robustness and measurement-induced percolation of the surface code}

\author{T.~Botzung}
\thanks{t.botzung@fz-juelich.de}
\affiliation{Institute for Quantum Information, RWTH Aachen University, D-52056 Aachen, Germany}

\affiliation{
Peter Gr\"unberg Institute, Theoretical Nanoelectronics, Forschungszentrum J\"ulich, D-52425 J\"ulich, Germany}

\author{M.~Buchhold}
\thanks{buchhold@thp.uni-koeln.de}
\affiliation{Institut  f\"ur Theoretische Physik, Universit\"at zu K\"oln, D-50937 Cologne, Germany}

\author{S.~Diehl}
\thanks{diehl@thp.uni-koeln.de}
\affiliation{Institut  f\"ur Theoretische Physik, Universit\"at zu K\"oln, D-50937 Cologne, Germany}

\author{M.~M\"uller}
\thanks{m.mueller@physik.rwth-aachen.de}
\affiliation{Institute for Quantum Information, RWTH Aachen University, D-52056 Aachen, Germany}

\affiliation{
Peter Gr\"unberg Institute, Theoretical Nanoelectronics, Forschungszentrum J\"ulich, D-52425 J\"ulich, Germany}

\date{\today}

\begin{abstract}
We examine the robustness of a logical qubit in the planar surface code subject to 'measurement-errors', i.e., to local Pauli measurements at known positions. This yields a measurement-only dynamics, which is driven by the competition between local Pauli measurements and stabilizer measurements. The setup differs from the conventional surface code for which errors are caused by decoherence and their positions are unknown. Yet, our setting allows us to examine the dynamics of the encoded logical qubit by using a combination of analytical arguments based on percolation theory and numerical simulations. Firstly, we show that for a single round of Pauli measurements only, the threshold corresponding to the irreversible loss of the logical qubit depends only on the rate of $
\hat X$- and $\hat Z$-measurements, and that this loss of logical quantum information is equivalent to the bond percolation transition in a 2D square lattice. Local $\hat Y$-measurements, which affect both $X$ and $Z$ stabilizers, erase the logical qubit only if all physical qubits along one of the diagonals are measured, and are thus negligible at large code distance.
Secondly, we examine the dynamics in the code by considering the interplay between rounds of Pauli measurements and rounds of stabilizer measurements. Focusing on the lifetime of the logical qubit, we obtain a rich phase diagram featuring different dynamical regimes. We argue that the loss of the logical qubit in this setting can still be understood by percolation theory and underpin our arguments with numerical simulations. 
\end{abstract}

\maketitle

\section{Introduction}

Measurements on many-body wave functions are fundamental for applications of quantum technology. For instance, they form a pillar of quantum state preparation and tomography, and importantly of quantum error correction~\cite{Nielsen2010Dec, Terhal2015Apr}. This puts the potential of measurements and the projective evolution of quantum states into the focus for current research. A prominent example, which recently attracted a lot of attention are measurement-induced phase transitions (MIPTs)~\cite{Li2018, Chan2019, Skinner2019, Li2019, Jian2020, Bao2020}. They appear in quantum systems evolved by frequent projective (or continuous) quantum measurements, and they manifest when the measurements are in competition with another generator of the dynamics. The latter can be, e.g., unitary circuits~\cite{Szyniszewski2019, Zabalo2020, Gullans2020Aug, Gullans2020Oct, Choi2020, Nahum2021, Sang2020Apr, Bao2021Feb, Lavasani2021Mar,Sierant2022Feb,Guoyi2022,Skinner2019,Li2018, Chan2019,Li2019,Jian2020,Fan2020,Bao2020,circuitreview,Romito2020,Morral2022,turkeshi2022measurement,Zabalo2022},
a unitary Hamiltonian evolution~\cite{Cao2019, Alberton2021,Buchhold2021,Botzung2021Nov, Piccitto2022Feb,Fuji2020Aug,Fuji2021Feb,Muller2022Jan, Block2022Jan,Zhang_2022,turkeshi2021measurementinduced,poboiko,Doggen2021,poboiko2023measurementinduced,Popperl,Chahine2023,tirrito2023}, open-system dynamics~\cite{Vovk2022Jun,Minoguchi2021,Ladewig}, or incommensurate measurements~\cite{Ippoliti2021,Lang2020Sep,Lavasani2022,Guoyi2023,Sriram2022,LiDecoding,Klocke_2022,Klocke2023,Negari2023}. In the latter case, the competition arises when the measured operators are selected from distinct, non-commuting sets. 

The dynamics at MIPTs has been linked to the working principle of quantum error correcting codes~\cite{Choi2020,Gullans2020Oct,Fan2021May,Li2021Mar,LiDecoding,Klocke_2022,Negari2023,Behrends2022}. In this picture local measurements take the role of adversaries, which continuously extract information from the system and eventually erase the logical quantum information encoded in a many-body wave function in terms of logical qubits. Once this happens, the wave function collapses onto a local product state, which lacks any particular features~\cite{Li2018,Li2019}. This loss of the logical information can be slowed down, when the local measurements are put in competition with a non-commuting generator of dynamics. Two complementary scenarios have emerged: (i) Unitary evolution generated by gates or a Hamiltonians leads to the scrambling of information in a (many-body) Hilbert space~\cite{Choi2020,Fan2021May, Hashizume2022Mar} and makes it thereby inaccessible to local measurements, yielding exponentially increased lifetimes of the scrambled qubits~\cite{Gullans2020Oct}. (ii) Alternatively, measuring stabilizer generators of the underlying quantum error correcting code, which anti-commute with the adversarial local measurements, continuously restores the computational subspace and thereby protects global operators of the encoded logical qubits from readout~\cite{LiDecoding,Lang2015Jul}.

Examining MIPTs may thus grant a complementary view on certain aspects of quantum error correction. Indeed, the MIPT in the measurement-only version of the one-dimensional quantum repetition code~\cite{Lang2020Sep,LiDecoding} signals the irreversible loss of the logical qubit, and it displays universal scaling inherited from two-dimensional bond percolation. In a similar spirit, measurement-induced dynamics inspired by quantum error correction have been studied in one- and two-dimensional systems, i.e., for the XZZX-code~\cite{Klocke_2022}, the toric code~\cite{Lavasani2021Dec,Negari2023} and the cluster model~\cite{Lavasani2021Mar}. While the loss of the logical information in these models corresponds to a topological phase transitions, its universal properties remain linked to $d+1$-dimensional percolation. 

One should note, however, that the analogy with the standard quantum error correction scenario is not complete: when doing measurements, the measured qubit as well as the measured operator (e.g. $\hat X, \hat Y, \hat Z$) are known to the experimenter. For realistic physical noise, however, both the position and the noise channel acting on a qubit are unknown. Instead, informed guesses for the locations and types of error processes that might have occurred need to be reconstructed from the syndrome information formed of the stabilizer measurement outcomes~\cite{PhysRevA.86.032324}. An exception is represented by \textit{qubit loss}, which is described by the quantum erasure channel, and for which the position of the lost qubit is assumed to be known~\cite{Grassl1997,Nielsen2010Dec}. Recently, techniques to detect and deterministically correct the loss of qubits in a quantum error correcting code have been demonstrated experimentally~\cite{Stricker2020Sep}. Thus, while MIPTs indicate the irretrievable loss of the logical information due to adversarial operations, they generally do \textit{not} indicate whether or not the information will be recovered by a quantum error correction protocol~\cite{LiDecoding}. For instance, the interpretation of the measured syndrome and proposed correction according to a given decoding strategy may mismatch the actual error pattern and lead to a logical error. The position of an MIPT thus yields an upper bound for the correctability of a quantum code, which may only be reached for qubit loss but not for general types of noise channels.

Nevertheless it is worthwhile to develop a deeper understanding of the connection between qubit losses, percolation theory, and MIPTs in paradigmatic quantum error correcting codes. Here, we focus on the planar 2D surface code, which is one of the leading contenders for scalable quantum error correction and fault-tolerant quantum computing, as highlighted by a series of recent experimental breakthroughs \cite{krinner2022realizing,zhao2022realization,google2023suppressing}. 

\begin{figure}[bt]
    \centering
    \includegraphics[width=1\columnwidth]{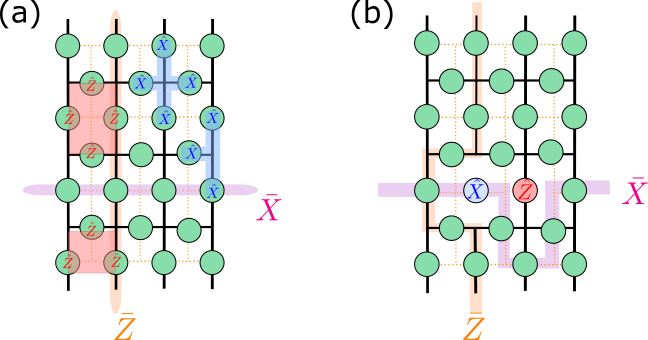}
    \caption{Planar surface code encoding one logical qubit. Physical qubits (green circles) are on the edges of a square lattice. (a) Examples of plaquette and star operators are depicted as red rectangles and blue crosses, respectively. Note the presence of weight-3 vertex stabilizer generators at the boundaries. The logical operator corresponds to a string operator $\bar{Z}$ connecting the upper and lower rough boundary of the lattice, whereas the logical $\bar{X}$ operators extends from the left to the right smooth boundary. (b) In the event of a local Pauli measurement, logical operators can be routed around measured qubits.}
    \label{fig: fig1}
\end{figure}

For sufficiently simple noise models, this code corresponds to an exactly solvable many-body model for quantum error correction. In particular, we analyze the robustness of an encoded logical qubit in a planar surface code (see Fig.~\ref{fig: fig1}(a)) subject to local Pauli measurements with known positions. While a sufficient number of Pauli measurements leads to the erasure of the logical information, its lifetime can be enhanced by stabilizer measurements. This competing measurement processes yield to a rich dynamics as a function of the respective measurement probabilities. 

We consider two different scenarios: (i) An initially prepared state in the code space and with one logical qubit subject to one round of local Pauli measurements $\hat X, \hat Y, \hat Z$. The measurements are performed with different probabilities $p_x, p_y, p_z$ and we examine the dependence of the survival probability of the logical qubit on the measurement probabilities. We show that the loss of logical quantum information can be mapped to a percolation problem for $\hat X$ and $\hat Z$ measurements. Furthermore, we find that in the thermodynamic limit the encoded logical information is robust against $\hat Y$ measurements. (ii) In the spirit of repeated rounds of quantum error correction cycles, we then modify the scenario: we perform multiple rounds of Pauli $\hat X, \hat Y, \hat Z$ measurements, where each round is followed by an additional round of stabilizer measurements. In this case, we argue that the loss of the logical qubit induced by $\hat X$ and $\hat Z$ measurements can be understood in a modified percolation picture and confirm this picture numerically. We show that the complexity of the percolation model increases with the number of stabilizer measurements. Finally, we observe a dynamical transition for the lifetime of the logical qubit in the unbiased case $p_x=p_y=p_z$. We find that the lifetime grows logarithmically in the system size for few stabilizer measurements and exponentially for many stabilizer measurements per round. 

The paper is organized as follows. In Sec.~\ref{sec: model}, we briefly introduce the measurement setup in the two-dimensional surface code and the observables we examine. In Sec.~\ref{sec: summary}, we provide a summary of the main results of this paper. In Sec.~\ref{sec: single_round}, we consider the case of a single round of a local Pauli measurement for $\hat X$ measurement only (Sec.~\ref{sec: X_Z_meas_one_round}), $\hat Y$ measurements (Sec.~\ref{sec: Y_meas}), and mixed Pauli measurements (Sec.~\ref{sec: XYZ_case}). In Sec.~\ref{sec: dynamics}, we discuss a discrete time evolution of the model when we add rounds of stabilizer measurements. In this setting,  we study several cases depending of the presence (or the absence) of certain types of Pauli measurements. In Sec.~\ref{sec: connection}, we focus on the case with only $\hat X$ measurements, and we show that the phase diagram can be understood within a picture based on percolation theory. In Sec.~\ref{sec: unbiased_dynamics}, we discuss the unbiased situation. We provide conclusions in Sec.~\ref{sec: conclusion}.

\section{Model}
\label{sec: model}

\subsection{Two-dimensional planar surface code}
We consider the two-dimensional planar surface code~\cite{Dennis2002Sep}, identified by the notation $[[n, k, d ]]$, with $n$ the number of physical qubits, $k$ the number of logical qubits, and $d$ the code distance. The latter is the smallest weight of a logical operator, corresponding to the minimal number of physical qubits that need to be manipulated in order to transform one code state into another. The code is defined by a set of stabilizer generators acting on the physical qubits which are located on the bonds of a square lattice (see  Fig~\ref{fig: fig1}). From now on, we refer to bonds as 'edges'. The stabilizer group is generated by plaquette $(\hat P)$ and star $(\hat S)$ operators 
\begin{equation}
\label{eq:stabilizer}
    \hat P_l =  \bigotimes_{i \in \text{plaquette } l} \hat Z_i,
   \ \ \ \ \hat S_l  =  \bigotimes_{i \in \text{star }l} \hat X_i,
\end{equation}
where $\hat Z_i$ and $\hat X_i$ are Pauli operators acting on qubit $i$ and stars and plaquettes are defined in Fig.~\ref{fig: fig1} (a). We use open boundary conditions, yielding weight-3 vertex stabilizer generators $\hat P_{l}$, and $\hat S_{l}$  at the  boundaries. This geometry leads to $N=n-1$ independent stabilizer generators, and therefore the planar surface code hosts one logical qubit ($k=1$), represented by one logical $\bar{Z}$ and one logical $\bar{X}$ operator. The code space $\{ \ket{C}\}$ hosting logical encoded states is defined as the simultaneous $+1$ eigenspace of all the stabilizer generators, i.e.
\begin{equation}\label{EqCode}
    \hat P_l \ket{C} = \hat S_l \ket{C}= \ket{C} \quad \forall \, \, l = 1, \ldots, N=n-1. 
\end{equation}

The logical operators correspond to a contiguous product of $\hat Z$ ($\hat X$) operators along strings of qubits connecting opposite rough (smooth) boundaries of the lattice as shown in  Fig~\ref{fig: fig1}(a). As required, each logical operator commutes with each stabilizer generator since the two operators always share an even number of qubits on the lattice. In addition, the logical generators anticommute, $\{\bar Z, \bar X\}=0$, since they intersect at a single or, more generally, an odd number of qubits. Acting with $\hat Z, \hat X$ on any state $|C\rangle$ from the code space (defined by Eq.~\eqref{EqCode}) leaves the state in the code space. The logical operators thus define a two-dimensionnal qubit subspace in which one logical qubit can be encoded. For states in the code space, the logical operators are, however, not unique. For instance $\hat P_l \bar X|C\rangle=\bar X\hat P_l |C\rangle \equiv \bar X |C\rangle$, for any plaquette $\hat P_l$. Thus $\bar X$ and $\bar X\hat P_l$ act identically on the state $|C\rangle$ (and the same for any star operator $\hat S_l$), which is depicted in Fig~\ref{fig: fig1}(b). Any two logical operators that can be mapped into one another by multiplication with stabilizer generators are thus equivalent. In the following, we will associate the case when both logical operators $\bar X, \bar Z$ are well-defined and $\{\bar X, \bar Z\}=0$ to the logical qubit being intact and well-defined.

\subsection{Projective measurements}
\label{sec: projectivemeasurements}
The survival of the logical qubits is challenged by local Pauli measurements and fostered by stabilizer measurements. This leads to a measurement-only evolution of the quantum mechanical wave function $|\psi\rangle$ corresponding to the initial state of the logical qubit. The dynamics is generated by projective measurements either drawn from the set of stabilizer generators $\hat O_l=\hat P_l, \hat S_l$ or from the set of local Pauli operators $\hat O_i=\hat X_i, \hat Y_i, \hat Z_i$. Each measurement evolves the wave function
\begin{equation}
    \label{eq: Pm}
|\psi\rangle\to \frac{1}{2\sqrt{p_\sigma}}(\hat{\mathds{1}}+\sigma\hat O_l)|\psi\rangle,
\end{equation}
depending on the measurement outcome $\sigma=\pm1$. The probability for each outcome is $p_{\sigma}=\tfrac12(1+\sigma \langle\psi|\hat O_l|\psi\rangle)$.

The measurements have two counteracting effects: (i) Measurements of local Pauli operators decouple physical qubit from all others, and the state is projected out of the code space. For instance, a local $\hat Z_i$ ($\hat X_i$) measurement acting on a qubit $i$ anticommutes with all the star (plaquette) and logical operators $\bar{X}$ ($\bar{Z})$, which include this qubit in their support. It thus makes the corresponding physical qubit temporarily inaccessible for the logical operators. If one or both of the logical operators are affected by the Pauli measurement, the operators can be routed around the measured qubit by multiplying them with one or several of the stabilizer generators. 
In Fig~\ref{fig: fig1}(b), we show the effect of two local Pauli measurements $\hat Z_i$ and $\hat X_{i+1}$. (ii) Measuring a stabilizer generator $\hat P_l$ ($\hat S_l$), which was previously affected by a Pauli measurement, adds the operator $\hat P_l$ ($\hat S_l$) back to the instantaneous stabilizer group, i.e., after the measurement we have $\hat P_l|\psi\rangle=|\psi\rangle$ ($\hat S_l|\psi\rangle=|\psi\rangle$)~\footnote{Without loss of generality we assume the measurement outcome $+1$.}. Thus, the measured stabilizer generator $\hat P_l$ ($\hat S_l$) can from this moment on be used again to route logical operators around inaccessible physical qubits. It is worth noticing that qubits belonging to the stabilizer generator  $\hat P_l$ are, after being measured, again partially accessible (only via $\hat P_l$). As long as the logical operators can be routed around inaccessible qubits by multiplication with well-defined stabilizer generators, the logical qubit and the initially encoded information are preserved. Conversely, the logical qubit is lost when there is no remaining path left for the logical operators to connect the two corresponding opposite smooth or rough boundaries, respectively. From this discussion, it is clear that measuring stabilizer generators acts as an error protection operation by reintegrating paths to route around the logical operators. However, it is essential to note that this error correction differs from the usual quantum error correction protocol, where the error syndrome formed of the stabilizer generator measurement outcomes is used in the decoding process to infer likely positions of errors and propose a suitable correction.

We organize the time evolution in measurement rounds. Each round is composed of (i) a series of local Pauli measurements, where on each qubit $i$ a local operator $\hat X_i, \hat Y_i, \hat Z_i$ is measured with probability $p_x, p_y, p_z$. This is followed by (ii) a series of stabilizer generator measurements, where each stabilizer generator $\hat P_l, \hat S_l$ is measured with probability $p_s$. We count time as the total number of rounds.

\subsection{Observables}
Our goal is to analyze the robustness of the logical operators $\bar X, \bar Z$ against measurements of local Pauli operators, in the presence (or absence) of stabilizer measurements. In this framework, local Pauli measurements act as errors, which eventually erase the logical operators. Stabilizer generator measurements protect the code space. In order to explore the dynamics, we perform numerical simulations of the measurement evolution. For this type of measurements and for a state that is initialized in the computational basis (i.e., an eigenstate of all stabilizer generators), the evolution can be efficiently implemented in the binary stabilizer formalism~\cite{Aaronson_2004}, see Appendix~\ref{ap: stab_binary}.

Once the evolution starts, the logical operators $\bar Z, \bar X$ will be deformed by the measurements. However, any information stored in the initial global logical qubit is preserved as long as both operators span a two-dimensional logical subspace, indicated by $\{\bar Z, \bar X\}=0$. The elimination of this subspace may happen in two possible ways: (i)  one of the logical operators (either $\bar Z$ or $\bar X$) is affected by a non-commuting measurement but can no longer be routed around inaccessible qubits by multiplication with instantaneously well-defined stabilizer generators, or (ii) both logical operators have been modified by the measurement evolution in such a way that they become ill-defined and effectively $[\bar X, \bar Z]=0$. In the latter case, the encoded qubit has been effectively projected onto a specific logical state (see App~\ref{ap: example_5code} for an example). In both cases, the logical operators do no longer define a single-qubit Hilbert space for the encoding of a general logical qubit state.

In the numerical simulations, we keep track of the instantaneous set of generators of the wave function $|\psi\rangle$ and of both logical operators $\bar X, \bar Z$. We consider the logical qubit to be lost when one of the above conditions (i) or (ii) is fulfilled for the first time. 
We introduce two complementary measures for the loss of the encoded information: we define (a) the logical success rate $\mathcal{R}$ as the probability that the logical information survives one round of Pauli errors and (b) the lifetime of the logical information $\tau$ as the average number of measurement rounds performed before the encoded information is lost.

\begin{figure*}[hbt!]
    \centering
    \includegraphics[width=1\textwidth]{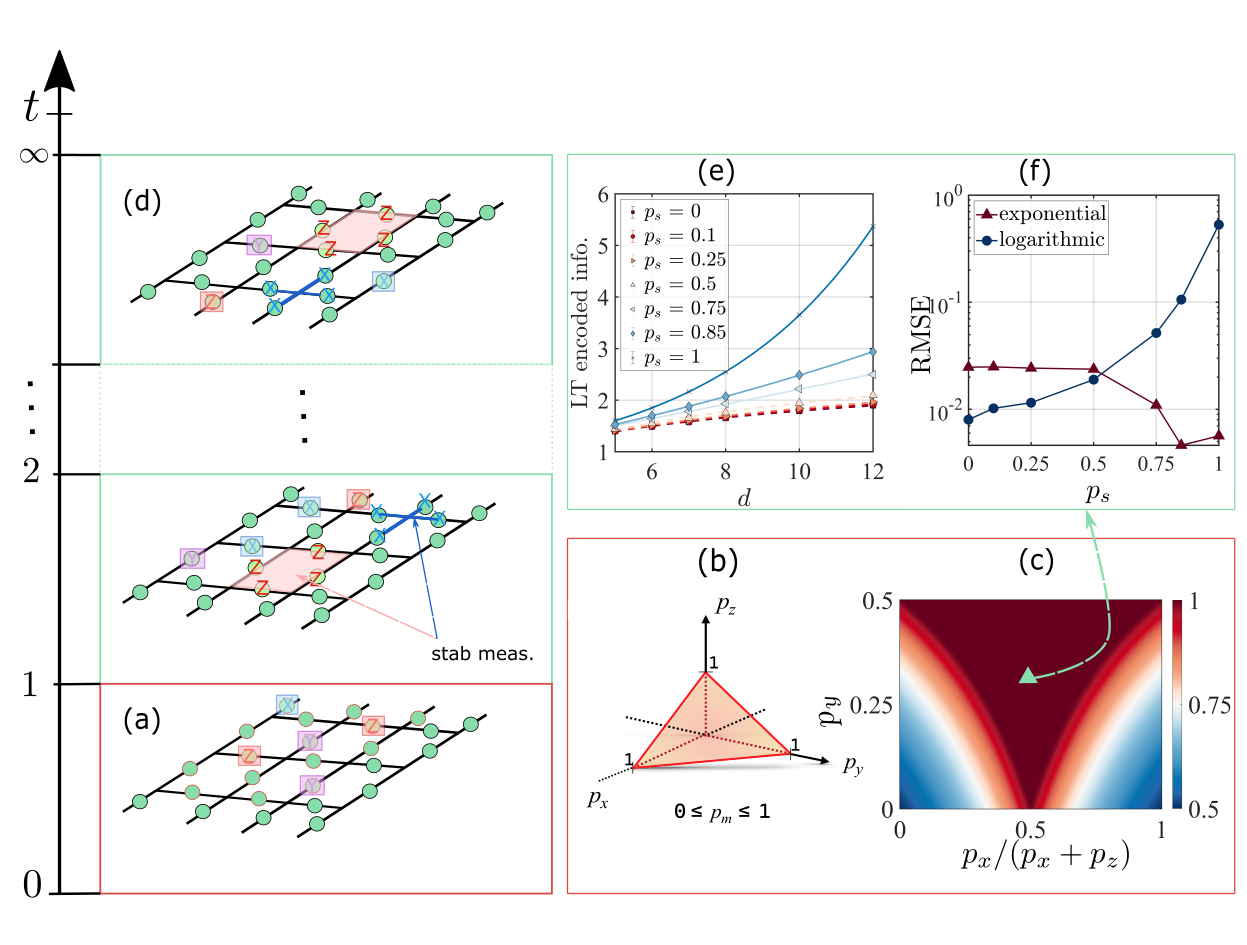}
    \caption{Projective evolution of the planar surface code, by measurement dynamics only. (a) The initial step comprises only local Pauli measurements, with the parameter space for the probability of measuring a qubit $p_m=p_x+p_y+p_z\le1$ in the $\hat X, \hat Y, \hat Z$ basis depicted as a hyperplane in (b). (c) Phase diagram of the survival probability of the encoded information as a function of the proportion of qubits measured in the $\hat X, \hat Z, \hat Y$-basis for the case of $p_x+p_y+p_z=1$. The green triangle indicates the unbiased situation where $p_x = p_y = p_z $. (d) Illustration of the discrete projective time evolution, where each round comprises a sequence of local Pauli measurements followed by a sequence of stabilizer measurements. (e) Finite-size scaling of the lifetime $\tau$ of the logical qubit with the distance of the code $d$ for the unbiased situation. The curves correspond to different probabilities of measuring stabilizers $p_s$ (see the color code), and $p_m = 0.95$. The solid lines are best fits of the form $a \exp(b d) + c$ and the dashed lines of the form $a \log(d) + b$, where $a, b, c$ are fit parameters and $d$ is the code distance. (f) The Root Mean Square Error (RMSE) for fits of the form  $a \log(d) + b$ (blue) and of the form  $a \exp(b d) + c$ (red) of the lifetime in (e). For $p_s \approx 0.5$, the exponential fits the data more accurately than the logarithm. In panel (e) dots are numerical data averaged over $10^{4}$ realizations.}
    \label{fig: summary}
\end{figure*}

\section{Summary of results}
\label{sec: summary}
We consider two different scenarios for the loss of the logical information: (A) a static one, where we stop the evolution after a single round of Pauli measurements and evaluate whether or not the logical information has survived, and (B) a dynamical one, where each round of Pauli errors is followed by a round of stabilizer measurements and we repeat this procedure until the logical information is lost. Both scenarios are illustrated in Fig.~\ref{fig: summary} and our main findings are summarized below.

\subsection{Single round of local Pauli measurements}
In the first scenario, in Sec.~\ref{sec: single_round}, we examine the robustness of the logical operators after a single round of local Pauli measurements, which are randomly drawn from the set $\{\hat X, \hat Y, \hat Z\}$. An example is illustrated in Fig.~\ref{fig: summary}(a). Each qubit is then measured at most once, with a total probability $p_m$. We label the probability for a $\hat X, \hat Y, \hat Z$-measurement as $p_{x}, p_y, p_z$, such that $p_m=p_x+p_y+p_z\le1$. This gives rise to a hyperplane as depicted in Fig.~\ref{fig: summary}(b). 

We distinguish three different cases for the measurement probabilities: (i) For $p_y=0$, i.e., in the absence of local $\hat Y$ measurements, the dynamics takes place on two decoupled sublattices (see Fig.~\ref{fig: fig1}). $\hat Z$-measurements commute with the $\bar Z$ logical operator and the plaquette operators but anti-commute with the $\bar X$ logical operator and the star operators. The situation is reversed for $\hat X$-measurements. Whether or not the $\bar X$ ($\bar Z$) logical operator remains well-defined thus depends only on the rate of $\hat Z$ ($\hat X$) measurements and both can be treated independently. The effect of $\hat Z$ measurements can be understood in a two-dimensional percolation picture: we consider a dual square lattice in which each physical qubit represents an edge and in which a $\hat Z$-measurement removes the corresponding edge from the lattice. The loss of the logical qubit occurs when a connected line of qubit measurements separates the lattice into two disconnected components along the $\bar X$- ($\bar Z$-) direction. The removal of edges in this two-dimensional (2D) lattice up to the point when there remains no path left which connects the two opposite boundaries corresponds exactly to the scenario of two-dimensional bond percolation on the square lattice. The threshold for the breakdown of percolation, i.e. the critical point of the transition, belongs to half of the edges being lost, i.e., a threshold of $p_z= 0.5$ ($p_x=0.5$)~\cite{DietrichStauffer2017Jan}. The analogy between the loss of the logical information in the surface code and percolation has previously been observed in the context of qubit loss~\cite{Stace2009May}, where a similar picture applies. 

(ii) For $\hat Y$ measurements only (i.e. $p_x=p_z=0$), the two sublattices for $\bar X, \bar Z$ are connected, which leads to a larger chance for the logical qubits to remain preserved. In this case the logical information is only eliminated if all the qubits on one of the two lattice diagonals are measured (see Fig.~\ref{fig: fig4}(a)). The number of qubits on both diagonals is $2(2d-1)$ and their fraction compared to the total number of qubits ($d^2 + (d-1)^{2}$) vanishes in the thermodynamic limit $d \to \infty$ as $\sim 4d^{-1}$. In the thermodynamic limit, the probability that all qubits on the diagonal are measured in a single round thus vanishes for any $p_y<1$. This implies a threshold of $p_m = p_y = 1$.

(iii) When all types of measurements are combined, the logical qubit is lost once the individual measurement probabilities $p_{x,y,z}$ exceed a threshold. We combine the observations of (i) and (ii), which motivates a threshold in the thermodynamic limit 
\begin{equation}
\label{eq: naive_xzy_1}
   \text{max}\{p_x,p_z\}=\frac{1}{2}.
\end{equation}
We validate this result with numerical simulations.
Thus, since $p_x+p_y+p_z\le1$, the robustness of the code against errors increases when the fraction of $\hat Y$ measurements (errors) becomes larger. The phase diagram for $p_x+p_y+p_z=1$ is shown in Fig.~\ref{fig: summary} (c)~\footnote{We note that a similar phase diagram has been obtained in the measurement-based quantum computation setup proposed in Ref.~\cite{Negari2023}: In this study, Pauli-measurements evolve the bulk of a 2D toric code ground state in order to generate entanglement at its one-dimensional spatial boundary. While the $\hat X$ or $\hat Z$ dominated regimes yield an area law entangled boundary, the $\hat Y$-measurement dominated regime induces a $\log^2 (L)$-growth of entanglement at a boundary of length $L$.}.

\subsection{Dynamics: competition between local Pauli and stabilizer measurements}\label{sec:summary dynamics}
In Sec.~\ref{sec: dynamics}, we consider a dynamical version of the setup. We examine the robustness of the logical operators for repeated rounds of random local measurements $\{\hat X, \hat Y, \hat Z\}$, which we denote as `error rounds.'  Each error round is followed by a round of random stabilizer generator measurements, in which star $\{\hat S_l\}$ and plaquette $\{\hat P_l\}$ operators are measured independently, each one with probability $p_s$. As discussed above, this second round can be seen as an error protection operation (Sec.~\ref{sec: projectivemeasurements}). This scenario is depicted in Fig.~\ref{fig: summary} (d). 

We characterize the dynamics by the lifetime of the encoded information $\tau(d)$, i.e., the lifetime of the logical operators $\bar X, \bar Z$, as a function of the code distance $d$. It quantifies the average number of rounds up to which the encoded information is preserved during the dynamics. The dependence of $\tau$ on the logical distance $d$ indicates whether or not the logical information is robust in the thermodynamic limit.

Here, we focus on two possible scenarios:\\
(i) Biased errors with $p_y=0$ but $p_x, p_z$ arbitrary, i.e., only $\hat X_i, \hat Z_i$ measurements occur. In this case, the dynamics is again separable: measurements of $\hat X_i$ operators and of plaquette stabilizers compete with each other but do not interfere with measurements of $\hat Z_i$ operators and star stabilizers, and vice versa. We thus focus on $\hat X_i$ and $\hat P_l$ measurements only. Examining the lifetime $\tau(d)$, we obtain the phase diagram as a function of $p_x$ and $p_s$. The scenario can be mapped to a modified 2D square graph, which hosts an effective percolation model based on the fraction of lost edges. The percolation model yields results that are consistent with the numerical simulations and it maps to conventional bond percolation in 2D in the limit $p_s = 1 $. The latter corresponds to a full restoration of the lattice at each stabilizer round, and thus to a scenario where the loss of the logical information can only occur within a single error round (see Sec.~\ref{sec: single_round}), as in scenario (A). The dynamics on the dual lattice composed of $\hat Z_i$ errors and $\hat S_l$ measurements is equivalent and independent for $p_y=0$.

(ii)  Further we consider totally unbiased errors, i.e. $p_x = p_y = p_z=p_m/3$. In this case, both $p_x,p_z<\frac{1}{2}$, are below the percolation threshold. Thus, eliminating the logical qubit with a single round of Pauli measurements will not happen due to the percolation mechanism. Instead, it would require a rare configuration of measured physical qubits, e.g., measuring $\hat Y$ for all qubits on one of the two diagonals of the surface code lattice, or performing an arbitrary Pauli measurement on each individual qubit. In order to understand the dynamical regimes for different values of the stabilizer measurement probability $p_s$, we assume here that each rare event has the same asymptotic dependence on the system size. We then focus on the latter, simple example, i.e., eliminating the logical qubit by measuring all physical qubits. 
In the limit $p_s=0$, i.e., when no stabilizer measurements are performed, the time-evolution of each physical qubit can be treated independently. Then the probability that at least one physical qubit remains unmeasured after $t$ time steps is $1-(1-(1-p_m)^t)^{d^2+(d-1)^2}$. For $p_m<1$, this yields a survival time, which grows logarithmically with the code distance  $\tau(d)\propto \log(d)$, see Sec.~\ref{sec: unbiased_dynamics}.
In the opposite limit $p_s=1$, i.e., when all stabilizer generators are measured in each round, the logical qubit can only be eliminated in a single round of Pauli measurement. Then we find that the scaling of the lifetime approaches an exponential law $\tau(d)\propto \exp(d)$, see Sec.~\ref{sec: unbiased_dynamics}. We can summarize both limits $p_s=0,1$ in the following way: For $p_s=0$ errors propagate uncorrected from round to round and different physical qubits become more and more uncorrelated due to the absence of stabilizer measurements. For $p_s=1$, all qubits are brought back to the computational basis after one stabilizer round and thus different measurement rounds are uncorrelated. We show that both regimes are robust and that the above discussed scaling behavior of the logical qubit life-time persists when tuning $p_s$ away from the limit $p_s=0,1$, see, e.g., Fig.~\ref{fig: summary}(e) for $p_m = 0.95$. Furthermore, an analysis of the numerically simulated lifetimes indicates that the logarithmic and exponential regime are separated by a phase transition, illustrated in Fig.~\ref{fig: summary}(f).

\begin{figure*}[bt]
    \centering
    \includegraphics[width=1\textwidth]{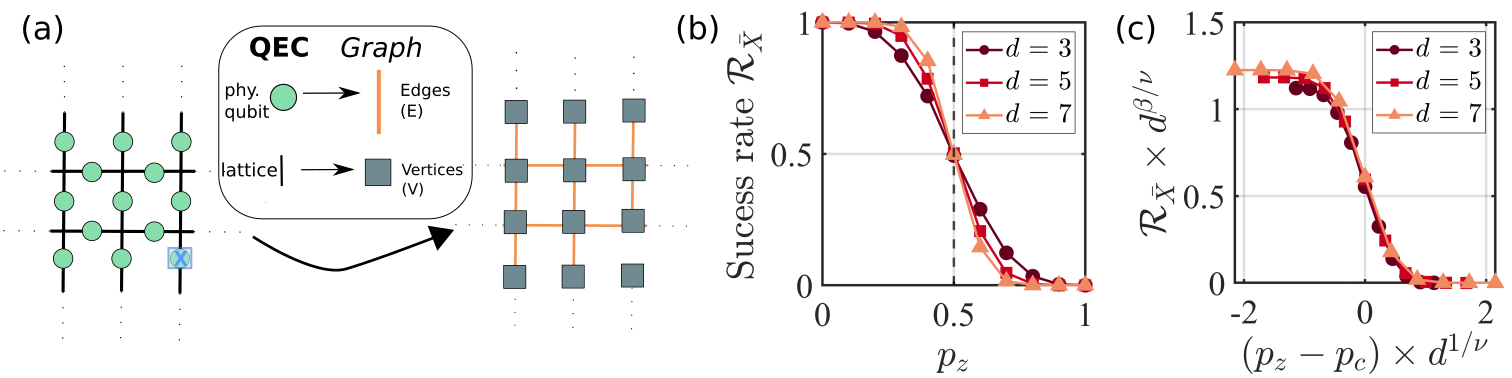}
    \caption{Single round of Pauli-$\hat X$ or -$\hat Z$ measurements. (a) Mapping the planar surface code into a graph for the $\hat Z$ lattice. The mapping is equivalent for the dual $\hat X$ lattice leading to two copies of an equivalent graph. The effect of local Pauli measurements $\hat X$ on the $\hat Z$ lattice and its corresponding graph is also shown. It corresponds to removing an edge in the graph. (b) Logical success rate of $\bar{X}$ ($\mathcal{R}_{\bar{X}}$) as a function of $p_Z$ for different distances of the surface code (see the color code). (c) Scaling function $\mathcal{R}_{\bar{X}} \times d^{\beta/\nu}$ as a function of $(p_z - p_c) d^{1/\nu}$ with imposed universal critical exponent of a two-dimensional bond percolation, $\beta=5/36$, $\nu = 4/3$, and a critical value for the transition $p_c=0.5$. The resulting collapse of the scaling is consistent with a two-dimensional bond percolation transition. In Appendix~\ref{ap: critical_1_round}, we extract a critical exponent $1/\nu \approx 0.73 \pm 0.03$, in agreement with the critical exponent for a 2D percolation transition. Numerical  data are obtained by averaging over $10^{4}$ realizations.}
    \label{fig: fig2}
\end{figure*}

\section{Single round of local Pauli measurements}
\label{sec: single_round}

We start by considering a single round of local Pauli measurements. This provides insights for the conditions under which the logical information is lost after local qubit measurements. 

\subsection{Single rounds of $\hat X$ and $  \hat Z$-measurement only}
\label{sec: X_Z_meas_one_round}

Consider a surface code with an initial state $|\psi\rangle$ in the code space, i.e., $\hat S_l|\psi\rangle=\hat P_l|\psi\rangle=|\psi\rangle$ and a logical operator $\bar Z$ or $\bar X $. The logical operators are initially defined as in Fig.~\ref{fig: fig1}(a). Measurements of local Pauli operators $\hat X_i$ ($\hat Z_i$) bring the state out of the code space by eliminating the operators $\hat P_l$ ($\hat S_l$) from the set of stabilizers. In addition, they deform the logical operators by routing them around measured qubits, see Fig.~\ref{fig: fig1}(b). Eventually, when measuring a sufficient number of Pauli $\hat X_i$ ($\hat Z_i$) operators, the logical operator $\bar Z$ ($\bar X$) can no longer be routed around the physical\ qubits, which have become inaccessible. Then the logical information initially stored in the state $|\psi\rangle$ is lost. In the language of the stabilizer formalism, this means that the logical operators $\bar Z$, $\bar X$ are replaced by a single Pauli operator. In the following we refer to this scenario also as the loss of the logical operators.

The Pauli measurements and the loss of the logical qubit can be described by a bond percolation model on the square lattice. In order to do so, we map the surface code to a graph $G(V, E)$. The graph contains edges $E$, representing the physical qubits, and vertices $V$, representing the lattice bonds, see Fig.~\ref{fig: fig2} (a). 

Measuring the operator $\hat X_i$ at edge $i$ eliminates this edge for the logical $\bar Z$ operator but leaves it unchanged for the logical $\bar X$ (vice versa a measurement of $\hat Z_i$ eliminates an edge for $\bar X$). In Fig.~\ref{fig: fig2}(a), we show the example of an $\hat X$ measurement (blue rectangle) and its effect on the corresponding graph. The  logical operator survives local qubit measurements as long as it is able to percolate through the lattice or, in other words, it is lost when the set of local Pauli measurements have caused a complete cut in the path of the logical operator. This loss of the logical information corresponds to a bond percolation transition on the graph $G(V,E)$. It occurs when the probability $p_E$ of eliminating an edge $E$ passes the value $p_E=0.5$. Since measurements of $\hat X$ ($\hat Z$) eliminate edges only for the logical operator $\bar Z$ ($\bar X$) both logical operators undergo separate and independent percolation transitions depending on the values of $p_x, p_z$. This scenario is reminiscent of surface codes in the presence of qubit loss. However, in the case of a qubit loss, since the qubit is completely traced out of the code space, it becomes  inaccessible for both lattices ~\cite{Stace2009May}. Then, a lost qubit removes the edge for both $\bar X, \bar Z$ and thus leads to a joint percolation transition for both logical operators. 

We confirm the percolation transition for $p_y=0$ by numerical simulations, which track the logical operator $\bar X$. Since $\bar X$ is unaffected by $\hat X$-measurements, its loss depends only on the probability $p_z$ to measure $\hat Z$. We compute the logical success rate $\mathcal{R}_{\bar{X}}$, i.e., the average probability that the logical operator $\bar X$ survives one round of local measurements. It is shown in Fig.~\ref{fig: fig2} (b) as a function of $p_z$, and for different distances of a planar surface code $d=3, 5, 7$, with a number of qubits $n = d^{2} + (d-1)^{2}$. The crossing of $\mathcal{R}_{\bar{X}}$ for different system sizes at $p_z = 0.5$ is consistent with the critical point of the percolation mapping. In Fig.~\ref{fig: fig2} (c), we further demonstrate that a scaling collapse of $\mathcal{R}_{\bar{X}}$ consistently reproduces the critical exponents of 2D bond percolation $\beta=5/36$ and $\nu =4/3$~\cite{DietrichStauffer2017Jan}. In addition, in Appendix~\ref{ap: critical_1_round}, we obtain the critical exponent $1/\nu \approx 0.73 \pm 0.03$ from a more reliable finite size scaling approach. It is in agreement with 2D percolation transition. The equivalent scenario is observed when tracking the logical operator $\bar Z$ for $\hat X$-measurements.

\subsection{Single rounds of $\hat Y$-measurements only}
\label{sec: Y_meas}

The measurements of Pauli $\hat Y$ operators interferes with both sublattices: $\hat Y$ operators anticommute with both plaquette and star stabilizers as well as the $\bar Z$ and the $\bar X$ logical operator. We show that this leads to an increased survival probability of the logical information.

\begin{figure*}[ht!]
    \centering
    \includegraphics[width=0.9\textwidth]{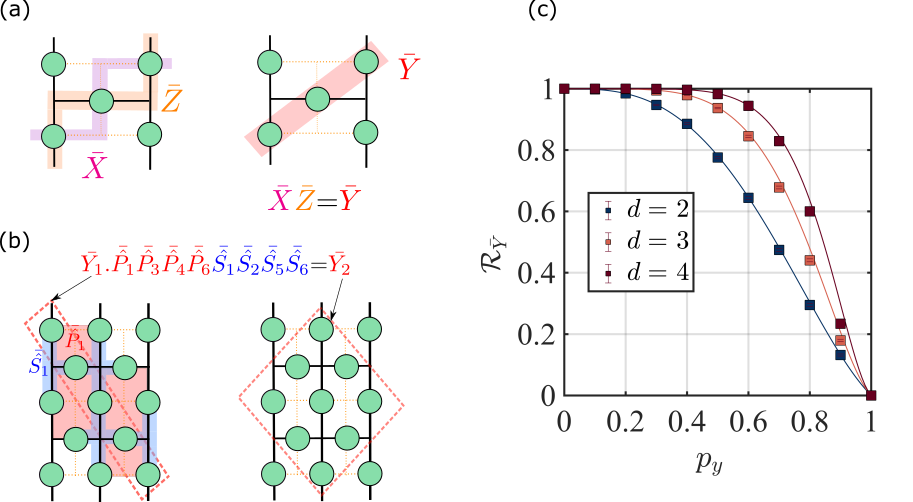}
    \caption{Single round of Pauli-$\hat Y$ measurements. (a) Geometrical construction of the $\hat Y$ logical operator $\bar{Y}$. On the left, by deforming the path of the $\bar{X}$ and $\bar{Z}$, we can create a product of $\hat Y$-operators that defines $\bar{Y}$. (b) Example of an alternative logical $\bar{Y}$-operator, which is obtained by multiplying the diagonal $\bar{Y}$-operator in (a) with plaquette and star operators. (c) Probability that the logical information survives a round of $\hat Y$ measurements as a function of $p_y$, and for different distances. The dots are the result of numerical simulations using the stabilizer formalism and a binary representation. We averaged over $10^4$ realizations. The lines are obtained from the analytical expression Eq.~\eqref{eq: psucessy}.}
    \label{fig: fig4}
\end{figure*}

We can rationalize this observation by considering a modified set of logical operators. Instead of using $\bar X$ and $\bar Z$, we can construct logical $\bar Y$ operators, which are composed of $\hat Y$ Pauli operators only. Two possible such $\bar Y$ operators for a $5$-qubit code are illustrated in Fig.~\ref{fig: fig4} (a) and (b). These are not affected by $\hat Y$-measurements unless each individual qubit of $\bar Y$ is measured. The remaining, anti-commuting logical operator, i.e., \textit{either} $\bar X$ \textit{or} $\bar Z$, can then be routed around inaccessible qubits by both plaquette and star operators. This yields an increased number of possibilities to circumvent the loss of the logical information, and thus a larger threshold for its survival. We show in Appendix~\ref{ap: example_5code} for an example of $5$ qubits that the destruction of the logical information requires indeed the measurement of all qubits contributing to $\bar Y$.

We assume now that this observation is general and holds for any $d>2$. Under this assumption, we derive a simplified threshold for $\hat Y$-measurements below. We then use numerical simulations to confirm that this threshold quite accurately captures the loss of the logical qubit. Qualitatively, we can understand the threshold by considering that the total number of qubits scales as $\sim d^2$, while the number of qubits on the diagonal grows linearly $\sim 2d-1$. By doing $\hat Y$-measurements randomly with a probability $p_y < 1$, the probability to measure all physical qubits on the diagonal converges to zero in the thermodynamic limit $d \to \infty$. This implies a threshold of $p_y=1$. Consequently, in order to eliminate the encoded logical information in this limit with randomly chosen $\hat Y$-measurements only, one has to measure essentially all physical qubits.

In order to obtain a more quantitative estimate, we start with the lowest number of measurements, which is required to eliminate $\bar{Y}$. This number corresponds to the number of physical qubits on the diagonal, which is given by $2d - 1$. The probability to measure these qubits is given by a binomial distribution $\mathcal{R}_{\rm{fail \, lowest  \, order }} = 2p_y^{2d-1} (1- p_y)^{n-(2d-1)}$. The factor $2$ is introduced to take into account the two diagonals. Extending the scenario to a larger number of measurements $k$, with $2d-1\le k\le d^2 + (d-1)^{2}$, we obtain ($n=d^2 + (d-1)^{2}$) 
\begin{equation}
\label{eq: pfaily}
    \mathcal{R}^{\rm{fail}}_{\bar{Y}}(n, p_y) = \sum_{k = 2d-1}^{n} \mathcal{C}(k, n) p_y^{k} (1- p_y)^{n-k}.
\end{equation}
 The binomial coefficients $\mathcal{C}(k, n)$ are 
\begin{equation}
\label{eq: coeff_Y}
    \mathcal{C}(k, n) = \left\lbrace 
    \begin{array}{cc}
                            & 2 \binom{n-(2d-1)}{k - (2d-1)},   \quad \forall k<4d-3 \\
                            \\
                            & 2\binom{n-(2d-1)}{k - (2d-1)} - \binom{n-(4d-3)}{k - (4d-3)},   \quad \forall k\geq4d-3.
    \end{array}
                            \right. 
\end{equation}
They count all possible combinations of measurements that lead to a failure of $\bar{Y}$. The pivotal value $4d-3$ is the total number of independent elements on the 2 diagonals ($4d-3 = 2(2d-1) -1$ (the diagonals have one common element)). When the number of measurements $k\geq 4d-3$, there is (at least) a common combination for the 2 diagonals. This implies that the total combination is lowered, as seen in Eq.~\eqref{eq: coeff_Y}. Using Eq.~\eqref{eq: pfaily}, the probability that the logical information survives in a surface code when subjected to $M^{Y}$ reads
\begin{equation}
\label{eq: psucessy}
    \mathcal{R}_{\bar{Y}} (n, p_y) = 1 - \mathcal{R}^{\rm{fail}}_{\bar{Y}}(n, p_y).
\end{equation}

In Fig.~\ref{fig: fig4} (c), we display $\mathcal{R}_{\bar{Y}}$ as a function of the probability $p_y$. We see that the numerical simulations (dots) for different distances (see color code) match the analytical predictions Eq.~\eqref{eq: psucessy}. Furthermore, we observe that the simulations for different distances cross at $p_y = 1$. This confirms that the threshold of $p_y=1$ for $\hat Y$-only measurements. In appendix~\ref{ap: example_5code}, we corroborate this statement by deriving the large $d$ limit of Eq.~\eqref{eq: pfaily}.

\subsection{Single round of $\hat X, \hat Y, \hat Z$ measurements}
\label{sec: XYZ_case}

\begin{figure}[ht]
    \centering
\includegraphics[width=0.9\columnwidth]{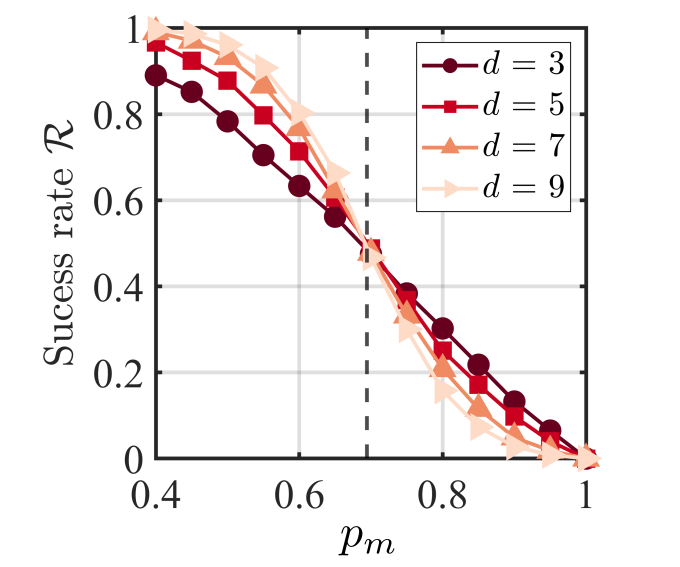}
    \caption{Single round of Pauli-$\hat X, \hat Y, \hat Z$ measurements. Probability of logical success as a function of $p_m=p_x+p_y+p_z$ for different distances. The dashed line indicates the analytical prediction from Eq.~\eqref{eq: naive_xzy}. Parameters are $(p_x,p_y,p_z) =(0.72,0.1,0.18)p_m$.}% \MB{This figure needs modification of the axis labels according to the text.  (a) Threshold for the total measurement probability $p_m$ obtained from Eq.~\eqref{eq: naive_xzy} as a function of $p_x, p_y, p_z$. (b) }}
    \label{fig: fig8}
\end{figure}

We now consider the most general case, in which operators $\hat X, \hat Y, \hat Z$ are measured with non-zero probabilities $p_{x,y,z}\ge0$. We exclude the scenario that a single qubit is measured by more than one Pauli operator. This yields the total probability that an arbitrary qubit is measured 
\begin{equation}
\label{eq: pm_XYZ}
\begin{split}
         p_m  & = p_x + p_z +  p_y.
\end{split}
\end{equation}
With the constraint $0 \leq p_m \leq 1$, we obtain a parameter space as depicted in Fig.~\ref{fig: summary} (b). 

In order to estimate the threshold in this situation, we combine the observations from the previous sections: we showed that $\hat X$- and $  \hat Z$-measurements eliminate the logical information  when either $p_x, p_z\ge0.5$ according to a percolation picture. Furthermore, for $\hat Y$-measurements, we found the threshold $p_y=1$. We thus conjecture that the fraction of qubits measured by $\hat Y$ will not impact the threshold and the percolation picture. This would yield the general threshold
\begin{equation}
\label{eq: naive_xzy}
\text{max}\{p_x, p_z\}\le0.5.
\end{equation}

In order to verify this result, we compare it with numerical simulations in Fig.~\ref{fig: fig8}.  We show the logical success rate $\mathcal{R}$ as a function of $p_m$ and for different distances $d=3, 5, 7, 9$. We fix $(p_x,p_y,p_z) =(0.72,0.1,0.18)p_m$. For $p_m=1$ this corresponds to a loss of the encoded information, since $p_x = 0.72 > 0.5$. The crossing between the curves takes place at   $p_m \approx 0.7$  (dashed line), which corresponds to $p_x*p_m=0.5$ as predicted from Eq.~\eqref{eq: naive_xzy}. This confirms that $\hat Y$-measurements do not observably modify the loss of the logical information.

\section{ Competition between local Pauli and stabilizer measurements}
\label{sec: dynamics}
We now extend the previous scenario and consider several rounds of measurements. As described in the beginning, each round is composed of a series of local Pauli measurements, followed by a series of stabilizer measurements, i.e., of plaquette and star operators. We initialize the system in the computational basis, i.e., with $+1$ eigenvalue of all stabilizers and with well defined logical operators $\bar Z, \bar X$. Then we perform
\begin{enumerate}
    \item  Pauli measurements with probability $p_m$;
    \item  stabilizer measurements with probability $p_s$;
    \item a check of the survival of $\bar{X}$, and $\bar{Z}$. 
\end{enumerate}
We denote time as the number of rounds that have been measured and we repeat the process until one (or both) of the logical operators are eliminated (see Fig.~\ref{fig: summary} (d)).

\subsection{Dynamics with Pauli-$\hat X$ and plaquette measurements.}

In order to study the dynamics in the presence of stabilizer measurements, we start with rounds of $\hat X$-measurements ($p_m=p_x$), each followed by a round of plaquette stabilizer measurements  (with probability $p_s$).
\subsubsection{Effective percolation theory}
\label{sec: connection}
The loss of the logical qubit in the presence of $\hat X$ (or $\hat Z$) measurements can be understood in a percolation framework by mapping the physical qubits to edges on a square lattice and interpreting the effect of a $\hat X$-measurement as removing the corresponding edge from the lattice. One would hope that this picture for the loss of the logical qubit remains applicable to the case when stabilizer measurements are added and several measurement rounds are performed. 

For small probabilities of Pauli measurements, this seems to be the case: Let us consider an initial state in the code space and then perform one measurement round, in which first one Pauli operator $\hat X_i$ and then one stabilizer $\hat P_l$ is measured, such that $\{\hat X_i, \hat P_l\}=0$. We understand that the effect of the $\hat X_i$-measurement is to remove the corresponding edge $i$ in the dual lattice. Then, since $\{\hat X_i, \hat P_l\}=0$ and since no other measurement has been performed, the stabilizer measurement will push the state back into the code space and thus, by definition, restore the previously removed edge. 

The evolution becomes, however, more complex, when the probability of Pauli measurements increases. To illustrate this, we consider a similar scenario where we start with an initial state in the code space and again perform one measurement round. Now we measure two different Pauli operators $\hat X_i, \hat X_{j}$ and then one stabilizer $\hat P_l$ is measured, such that $\{\hat X_i, \hat P_l\}=\{\hat X_j, \hat P_l\}=0$, i.e., the stabilizer is incompatible with both Pauli operators. In the dual lattice, this will first remove the edges $i$ and $j$. But then the measurement of $\hat P_l$ will restore neither of them completely. The measurement of $\hat P_l$ restores one qubit of information but this is not sufficient to restore either edge $i$ or $j$, instead it provides information on the shared parity between the edges. This creates a new edge, which is shared between the two original ones: if further a stabilizer $\hat P_m$ is measured, which anticommutes with either $\hat X_i$ or $X_j$, both edges will be reintegrated in the dual lattice simultaneously. 

This example illustrates that the bond percolation picture needs to be modified when performing several measurement rounds and including stabilizer measurements. We show in Appendix~\ref{ap: validity_square} that for this scenario the planar code cannot be mapped to a simple graph since additional complexity emerges from stabilizer measurements, which do not only restore lost edges in the original square lattice but may add new connections between different nodes.

At the same time, one expects the dynamics not to become too complicated either: each Pauli measurement makes one physical qubit inaccessible for the logical operators, while each stabilizer measurement restores one qubit of information, which is shared by several physical qubits and which commutes with both logical operators. This qubit, local or not, can be used to route the logical operators around inaccessible qubits, i.e., as an effective edge (local or not) in the dual lattice. In the stationary state, which is reached after several rounds of measurements, we thus expect a stationary distribution of edges, which may connect nodes also over larger distances. The logical qubit is preserved if the edges form a connection between the two opposite sides of the lattice and is lost otherwise. Despite the potentially modified geometry of edges, this remains the same phenomenology as in the original percolation framework. 

In order to obtain an estimate for the dynamics with stabilizer measurements, we thus consider an effective phenomenological percolation model: (i) Single Pauli measurements happen with probability $p_x$ and remove an accessible edge and make it inaccessible. (ii) Stabilizer measurements reintegrate inaccessible edges back into the graph with an effective probability $\tilde p_s$. This considers scenarios in which more than one stabilizer measurement is required in order to reintegrate a particular edge. It can be expressed as
\begin{equation}
\label{eq: ansatz}
    \tilde{p}_{s} = \sum_{k=1}^{n} \alpha(k) p_{s}^{k}.
\end{equation}
Here, $\alpha(k)$ can be seen as the probability that a particular edge requires $k$ plaquette measurements, each occurring with probability $p_s$, in order to be reintegrated. For a given set of measurements and a given edge, the number of necessary plaquette measurements $k_E$ is known, i.e., $\alpha(k)=\delta_{k,k_{E}}$. Here, we replace $\alpha(k)$ by its statistical average over many edges and measurements. In any case, it is normalized $\sum_k \alpha(k)=1$. In the thermodynamic limit, we find that $k\le3$ yields already accurate results.

\begin{figure*}[hbt]
    \centering
    \includegraphics[scale=0.68]{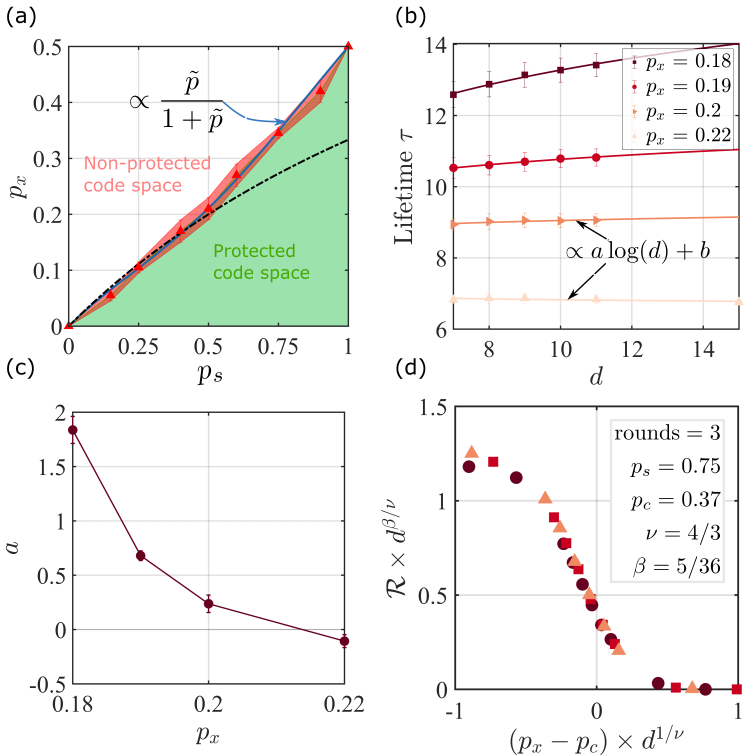}
   \caption{Dynamics of biased errors with only $\hat X$ and $\hat P_l$ measurements.  (a) Phase diagram of the correctability of the code as a function of $p_x$ and $p_{s}$. Triangles are numerical extrapolation from the lifetime $\tau$. The black dashed line is the dilute  regime solution Eq.~\eqref{eq: dilute_solution}.
   The dashed blue line is a fit of the form of Eq.~\eqref{eq: thres_x_only}, where we have used the ansatz for $\tilde{p}_{s}$ Eq.\eqref{eq: ansatz} up to the third order.  (b) Lifetime $\tau$ of the encoded logical information as a function of the code distance $d$ for different $p_x$, and a fixed $p_{s} = 0.4$. Dots are numerical data averaged over $5\times10^4$ realizations, and lines are best fits of the form $a\log(x) + b$. The dotted lines are fits with a constant value, and are a guide to the eye. (c) Extrapolated coefficient $a$ from the fit in (b) as a function of $p_x$. (d) Finite size scaling of the success rate $\mathcal{R}$ after $t=3$ rounds of measurements and for $p_s=0.75$. The scaling collapse as a function of $(p_x -p_c)$ is consistent with the critical exponents of the 2D bond percolation transition $\nu=4/3$ and $\beta=5/36$.}
    \label{fig: dynamics_result}
\end{figure*}

The percolation transition, associated with the loss of the logical information, occurs when more than half of the edges have become inaccessible. We thus study the fraction of the lost edges at time $t$, defined as 
\begin{equation}
    \label{eq: frac_lost}
\mathcal{F}(t)  = \frac{\text{number of lost edges at time t}}{\text{total number of edges}}.
\end{equation}
The dynamics of the effective model with the probabilities $\tilde p_s$ yields ($d\mathcal{F}(t)=\mathcal{F}(t)-\mathcal{F}(t-1)$)
\begin{align}
d\mathcal{F}(t)  &= \underbrace{p_x[1 - \mathcal{F}(t-1)]}_{\textrm{\textcolor{cyan}{loss}}}- \underbrace{\tilde{p}_{s} [\mathcal{F}(t-1) - p_x \mathcal{F}(t-1)]}_{\textrm{\textcolor{teal}{recovery}}}\nonumber\\
&=\mathcal{F}(t-1) \big[  p_x \tilde{p}_{s}  - p_x - \tilde{p}_{\rm p} \big] + p_x.
\label{eq: dynamic_fraction}
\end{align}
The nonlinear term $\sim \tilde p_sp_x$ occurs from processes where Pauli measurements eliminate qubits that have been restored in the same round. For a fixed value of $p_s$ and $\mathcal{F}(0)=0$ the time-dependent fraction is
\begin{equation}
\mathcal{F}(t) =  p_x  \big( 1 - \frac{1}{\kappa} \big) \exp(-\kappa t)   + \frac{p_x}{\kappa},
\label{eq: fraction_solution}
\end{equation}
with $\kappa = [p_x+\tilde p_s-p_x \tilde{p}_{s}  ]>0$. The fraction of lost edges reaches the stationary value $\mathcal{F}(t \to \infty)= \frac{p_x}{p_x + \tilde{p}_{s} - p_x \tilde{p}_{s}}$. From the percolation condition $\mathcal{F}<0.5$, this yields readily the threshold
\begin{equation}
\label{eq: thres_x_only}
\begin{array}{ccc}
    p_x & \le & \frac{\tilde{p}_{s}}{1 + \tilde{p}_{s}} \\
\end{array}
\end{equation}
for a yet to determine re-integration probability $\tilde p_s$. 

\subsubsection{Effective reintegration probability}
The effective probability $\tilde p_s$ is bounded from above by $\tilde p_s\le p_s$, which corresponds to the case that an inaccessible edge requires only a single stabilizer measurement in order to be reintegrated to the graph. This picture is true if the inaccessible edge is surrounded only by active ones, i.e., for low probabilities $p_x$ and large probabilities $p_s$. In  Appendix~\ref{ap: validity_square}, we confirm that  $\tilde{p}_{s} =p_{s}$ indeed holds for $p_x\le0.15 p_s$. This condition, however, is fulfilled only away from the percolation transition. 

An improved estimate can be obtained from a local density or mean-field type approximation: A local region of $k$ inaccessible edges requires exactly $k$ stabilizer measurements in order to reintegrate each edge in the graph. An area $A$ with a fraction $\mathcal{F}$ of lost edges thus needs an average number of $A\times\mathcal{F}$ stabilizer measurements to reintegrate each edge back into the graph. This simple picture suggests an effective reintegration rate per missing edge of 
\begin{equation}
\label{eq: localapp}
    \tilde{p}_{s} = (1-\mathcal{F}(t-1)) p_s
\end{equation}
 at time $t$. This simple estimate reproduces the limit $p_x\ll p_s$ and at the percolation threshold ($\mathcal{F}=0.5$) it yields
\begin{equation}
\label{eq: dilute_solution}
    p_x = \frac{p_s}{2 + p_{s}}.
\end{equation}
Indeed, numerical simulations confirm that this threshold is a good estimate in the regime $p_s\le0.5$, where correlations between different stabilizer measurements are rare and can be treated on a mean-field level (see Fig.~\ref{fig: dynamics_result}(a)).

\subsubsection{Finite size scaling of the success rate}

In order to compare this prediction with numerical simulations, we compute the lifetime of the encoded logical information $\tau$ in the planar code. The scaling of $\tau$ is a sensible indicator of the transition: it increases (decreases) with the system size when the local measurement probability is below (above) the actual threshold. In Fig.~\ref{fig: dynamics_result}(b), we show the lifetime $\tau$ as a function of the code distance $d$ for different $p_{x}$, and a fixed $p_s =0.4$. The dots are numerical data, and the lines are best fits to the form $a \log(d) + b$. The failure of the code is then indicated by a sign change of $a$ as a function of $p_x$, e.g., as shown in Fig.~\ref{fig: dynamics_result}(c). The critical value of $p_x$, i.e., the threshold, is located at $a=0$. For instance, we find $p_x^{c} \approx 0.17$, for $p_s =0.4$. The resulting phase diagram is displayed in Fig.~\ref{fig: dynamics_result}(a). The red triangles are numerical estimates, and the blue dashed line is the threshold Eq.~\eqref{eq: thres_x_only}, where we have used the ansatz Eq.~\eqref{eq: ansatz} up to the third order with the fitted coefficients $(\alpha(1),\alpha(2),\alpha(3))= (0.5,-0.375,0.875)$.

In order to further corroborate that the loss of the logical qubit during several measurement rounds corresponds to a bond percolation transition, we analyze the scaling behavior of the success rate $\mathcal{R}$ after a fixed number of rounds $t$.  In Figure~\ref{fig: dynamics_result}(d), we show the finite size scaling of $\mathcal{R} d^{\beta/\nu}$ as a function of $(p_x -p_c) d^{1/\nu}$ for $p_s=0.75$ and $t=3$ rounds of  measurements. A convincing scaling collapse is obtained for the  critical exponents $\nu =4/3$ and $\beta=5/36$, which correspond to 2D bond percolation. The location of the critical point  $p_c=0.37$ is extracted from the crossing point of $\mathcal{R}$ for different sizes $d$.

\subsubsection{Modifications for $\hat Z$- or $\hat Y$-only measurements}
As discussed in the previous section, measurements of Pauli-$\hat Z_i$ operators and star stabilizers $\hat S_l$ do not interfere with plaquette- and Pauli-$\hat X_i$ measurements and lead to an equivalent dynamics in the percolation framework. 

The case of performing only $\hat Y$-measurements, however, is different. Here, again, the logical information is lost only when each qubit of a global $\bar Y$ logical operator, e.g., one of the diagonals of the code has been measured (see Eq.~\eqref{eq: fraction_solution}). The loss of the logical qubit in finite time $t<\infty$ then again yields a threshold of $p_y=1$. We corroborate this statement in Appendix~\ref{ap: threshold_y_dynamics}.

\subsection{Dynamics with unbiased $\hat X, \hat Y, \hat Z$ measurements}
\label{sec: unbiased_dynamics}

\begin{figure}[hbt]
    \centering
    \includegraphics[width=1\columnwidth]{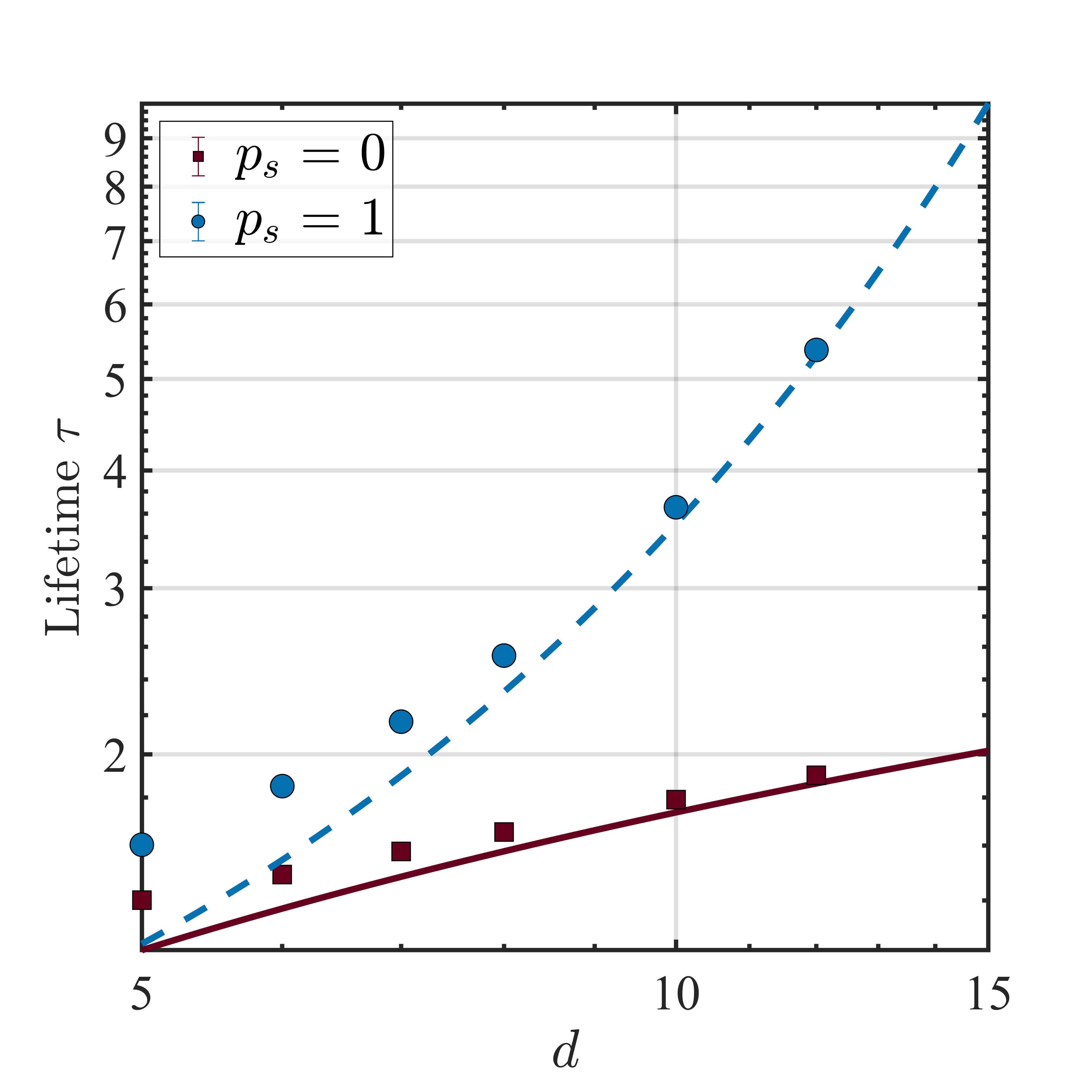}
   \caption{Lifetime $\tau$ of the encoded information as a function of the code distance $d$ for $p_s=0$ (blue) and $p_s = 1$ (red). the lines are proportional to Eqs.~\eqref{eq: survival_ps0} (red solid line) and \eqref{eq: survival_ps1} (blue dashed line). Dots are numerical data averaged over $10^{4}$ realizations.}
    \label{fig: unbiased_typical_time}
\end{figure}

Finally, we consider again the case of unbiased errors, i.e., for $p_x = p_y = p_z=p_m/3$. As we discussed in Sec.~\ref{sec:summary dynamics}, this set of parameters is away from the percolation transition. Regardless of the value of $p_m$, the probability $p_x, p_z < \frac{1}{2}$. The logical qubit is then generally robust and may only be lost due to a rare measurement configuration, e.g., measuring one of the diagonals with $\hat Y$-measurements or measuring all qubits at once with arbitrary Pauli operators. Since such rare events vanish in the thermodynamic limit, the logical qubit survives at any stabilizer measurement probability $0 \leq p_{s} \leq 1$. To show this explicitly, we propose the analytical scaling of the survival time of the encoded information on the two limits: (i) without stabilizer measurement ($p_s=0$), and (ii) with a complete restoration of the lattice at each step ($p_s=1$). In both cases, the survival time grows with the code distance $d$, which we validate with numerical simulations. 

In order to derive the analytical scaling of the lifetime, we  consider the particular rare event that all physical qubits are measured in finite time as the most likely event in the thermodynamic limit. For a finite system $d<\infty$, this means that one has to measure all the physical qubits in $t$ steps.
We consider the two limiting cases (i) without stabilizer measurement ($p_s =0$) and (ii) with a complete restoration of the lattice after each round of Pauli measurements ($p_s = 1$). 

(i) In the case without stabilizer measurements, we can treat each individual qubit independently. Hence, the probability of measuring a qubit in one time step equals $p_m$. This implies that the probability for a given qubit not being measured after $t$ steps is $(1-p_m)^t$ and that the probability of this qubit to be measured at least once is $1-(1-p_m)^t=1-e^{-\eta t}$, where $\eta=-\log(1-p_m)$. Now we ask what is the probability for $n$ qubits that at least one qubit has not been measured after $t$ steps. This is exactly the survival probability $1-(1-e^{-\eta t})^n$. Drawing the limit of large times $\eta t\gg1$ and a large number of physical qubits $n\gg 1$, the survival probability becomes $1-(1-e^{-\eta t})^n\approx 1-e^{-ne^{-\eta t}}$. The lifetime $\tau$ (or typical survival time) indicates the time after which the probability has dropped below $1/e$. It is obtained by equating the exponential $ne^{-\eta \tau}=1$, yielding \begin{equation}
\label{eq: survival_ps0}
    \tau \sim -\frac{\log(d^2+(d-1)^2)}{\log{(1-p_m)}},
\end{equation}
where we used that $n=d^2+(d-1)^2$ is the number of physical qubits .

(ii) Let us now consider the opposite limit $p_s=1$. This means that the state is brought back to the code space in each individual measurement round. Therefore the logical qubit needs to get lost by a rare event in one of the $t$ steps. The most likely rare measurement configuration is then the one where all qubits on the diagonal are measured by a $\hat Y$ operator (please note that this event is less likely in the previous case (i) since a single $\hat Z, \hat X$ measurement on the diagonal excludes this scenario to happen at any later $t$). The probability for this event is $p_y^{4d-2}$. Hence, the probability for the logical information to survive after $t$ steps is $(1-p_y^{4d-2})^t\approx e^{-tp_y^{4d-2}}$, where the first term approaches the second if $n$ is large. This yields the scaling of the lifetime of the logical qubit \begin{equation}
\label{eq: survival_ps1}
    \tau \sim p_y^{4d-2}.
\end{equation} 

In Fig~\ref{fig: unbiased_typical_time}, we show the lifetime $\tau$ as a function of the code distance $d$ for the case $p_s=0$ and $p_s=1$. The dots are the numerical simulations, and the lines are proportional to the Eq.~\eqref{eq: survival_ps0} (red solid line) and \eqref{eq: survival_ps1} (blue dashed line). In both cases, the scaling formulas predict correctly the leading order in the large $d$ limit. For small code distances, we observe a discrepancy that we attribute to finite-size corrections. In Fig.~\ref{fig: summary} (e), we show the lifetime $\tau$ as a function of $d$ for different values of $p_s$. We show that both regimes are robust away from the limit $p_s=0, 1$. By extracting the Root Mean Square Error (RMSE) for fits of the form  $a \log(d) + b$ (blue) and of the form  $a \exp(b x) + c$ (red) of the lifetime in Fig.~\ref{fig: summary} (e), we show that that the logarithmic and exponential regimes are separated by a sharp transition, illustrated in Fig.~\ref{fig: summary}(f). 

\section{Conclusion}

\label{sec: conclusion}

In this work, we have studied the loss and the survival of a logical qubit encoded in a planar surface code in the presence of local Pauli-measurements and measurements of stabilizer generators.

First, we considered a single round of only local Pauli measurements. Without $\hat Y$-measurements, i.e., when measuring only Pauli-$X$ Pauli-$Z$ operators, a robust phase where the logical qubit is preserved (but scrambled) is separated from a phase in which the logical qubit is irremediably lost. The transition between the two phases can be understood in a 2D bond percolation framework, and is reminiscent of the scenario of qubit losses. In contrast to $\hat X$- and $\hat Z$-measurements, the measurement of Pauli-$Y$ operators has a less severe effect: the loss of the logical qubit requires rare spatial configurations of $\hat Y$ measurements, the most likely one of which being the measurement of all physical qubits along the diagonal of the code. In the thermodynamic limit, such rare situations become increasingly unlikely, making the logical qubit asympotically perfectly robust against $\hat Y$-measurements. 

The so-established framework, (i) a percolation-type mechanism for the loss of the logical qubit due to $\hat X$- and $\hat Z$-measurements and (ii) the loss due to rare measurement configurations for $\hat Y$-measurements, persists in a dynamical framework, where each sequence of Pauli-measurements is followed by a sequence of stabilizer measurements. We motivated this picture with a set of analytical approaches suited to each scenario, i.e., an effective percolation theory and a rare event description, and confirmed the results with numerical simulations. 

This establishes a general and complete picture for the loss of the logical qubit due to local Pauli-measurements in the surface code. We highlight the role of percolation as the leading mechanism that persists in the thermodynamic limit. The latter also shows an interesting relation to the  qubit loss. Indeed, the qubit loss phenomenon can be viewed as a measurement of a physical qubit (at a given position) with an unknown outcome. However, this differs from local Pauli-measurements, where the measurement outcome is also known. In this work, we found that for a planar surface code, some specific Pauli measurements lead to a percolation transition identical to the one observed in the context of a surface code with qubit loss. It would be worth investigating in more depth the connection between these two processes. For instance, it would be interesting to determine and compare the threshold for the loss of the logical qubit in higher dimensional codes in the presence of Pauli measurement or qubit loss.  Appealing situations are given by topological color codes, for which optimal error correction protocols and fundamental thresholds are often not known. In light of our findings, it would also be interesting to extend our study to measurement dynamics beyond Clifford dynamics, e.g.~by allowing single-qubit measurements about arbitrary axes, or by allowing for weak measurements of single qubits and stabilizer operators.

\begin{acknowledgments}
TB and MM acknowledge support by the European Research Council (ERC) via ERC Starting Grant QNets Grant Number 804247. MM acknowledges support by the EU H2020-FETFLAG-2018-03 under Grant Agreement number 820495, and by the Deutsche Forschungsgemeinschaft through Grant No. 449905436. MB and SD acknowledge support by the  Deutsche Forschungsgemeinschaft (DFG, German Research Foundation) under Germany’s Excellence Strategy Cluster of Excellence Matter and Light for Quantum Computing (ML4Q) EXC 2004/1 390534769 and by the DFG Collaborative Research Center (CRC) 183 Project No. 277101999. The authors gratefully acknowledge the
computing time provided to them at the NHR Center
NHR4CES at RWTH Aachen University (project number p0020074).
\end{acknowledgments}
%- project B02, and by the European Research Council (ERC) under the Horizon 2020 research and innovation program, Grant Agreement No. 647434 (DOQS).

\appendix

\section{Stabilizer formalism:  measurements in the binary representation}
\label{ap: stab_binary}
\subsection{Formulation of measurements}
\label{ap: effect_meas}
Here, we briefly summarize the effect of measurements in the binary stabilizer representation~\cite{LiDecoding,Aaronson_2004}. Without loss of generality we adopt the conventional notation and consider a state with stabilizer generators $s_i$, and logical operators $\bar{L}$. The measurement of an operator $g$ leads to one of the possible scenarios below:
\begin{list}{$\square$}{\leftmargin=0em \itemindent=1em}
\item $g$ commutes with all the generators $s_i$ and logical operators. In this case, the measurement $g$ does not affect the state. 
\item $g$ anti-commutes with one or more generators but commutes with the logical operators. In this case, the state of the system is obtained by choosing one of the anti-commuting generators $s_i$ and replacing it by $g$. All other anti-commuting generators $s_k$ are replaced by the product $s_i s_k$. As a consequence, $[ s_i s_k, g ] = 0$. Since $[\bar{L}, g] =0$, the logical information is not modified by the measurement.
\item $g$ anti-commutes with one or more generators and with one or more logical operator $\bar{L}$. This case is similar to the previous one. The significant difference is that in addition we modify the logical operators also by multiplying them with $s_i$, i.e.,  $\bar{L}\to s_j \bar{L}$ (see Fig~\ref{fig: fig1}(b) for an illustrative example). The logical information is preserved.
\item $g$ commutes with all generators $s_i$ but anti-commutes with one or more logical operators. Since there is no generator that anti-commutes with $g$, the logical operator is replaced by $g$ and thereby eliminated. The logical information is lost.
\end{list}

\subsection{Binary representation of a stabilizer code}

\begin{figure}[hbt]
\label{fig: check_matrix}
    \centering
    \includegraphics[width=1\columnwidth]{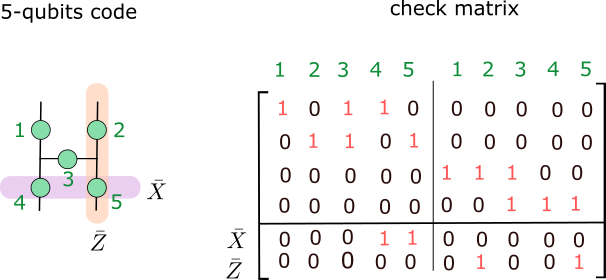}
    \caption{Planar code with $5$ qubits and its stabilizer (check) matrix.}
\end{figure}

We use the conventional binary representation of the stabilizer formalism~\cite{Aaronson_2004} to simulate the dynamics of the surface code under measurements. In this representation (we drop operator hats for convenience), any Pauli string $g$ over $N$ qubits is, up to a global phase, uniquely expressed by a $1 \times 2N$ vector $w=(u,v)$ of binary entries $0,1$, via
\begin{equation}
    g = e^{i\theta} \Pi_{i=1}^{N} X_i^{u_i} \Pi_{i=1}^{N} Z_i^{v_i}.
\end{equation}
Hence, an element $g$ of the Pauli group can be represented via a $2 \times N$ row vector, which we denote from now on as $r(g)$. Some useful properties result from this representation. For instance, if we define a matrix $2N \times  2N$ as
\begin{equation}
    \Lambda = \begin{bmatrix} 0_{N \times N}  & \mathds{1}_{N \times N} \\
        \mathds{1}_{N \times N}   &    0_{N \times N}
    \end{bmatrix},
\end{equation}
one can see that two elements of the Pauli group $g$ and $g'$ commute if and only if $r(g)\Lambda r(g') = 0$. In addition, the multiplication between $g$ and $g'$ ($gg'$) corresponds simply to $r(g) + r(g')$.  Note that in this representation, the arithmetic is done modulo two. 

This provides a practical way to represent the stabilizer generators of a code $s=\langle g_1, g_2 ..., g_N \rangle $,  the logical operators $\bar{X}, \bar{Z}$ as well as the measurement operations.
In Fig.~\ref{fig: check_matrix}, we present an example of a stabilizer matrix for the planar distance-2 surface code formed of 5 qubits.

\section{$\hat Y$ measurement only}

\subsection{Example of $\hat Y$ measurements in the 5-qubit surface code}
\label{ap: example_5code}
Here, we provide an example for $\hat Y$-only measurements in the 5-qubit surface code. It illustrates how a sequence of $\hat Y$-measurements eliminates the encoded logical qubit. We explicitly show that a measurement-sequence corresponding to a diagonal of the code ($\bar{Y}$) is required to eliminate the encoded information. 

We start with the stabilizer set and logical operators (we drop operator hats for convenience)
\begin{equation}
\begin{array}{ll}
    S & =  \langle Z_1 Z_2 Z_3, X_1 X_3 X_4, Z_3 Z_4 Z_5, X_2 X_3 X_5  \rangle, \\
    \bar{X} & = X_4 X_5, \\ 
    \bar{Z} & = Z_2 Z_5.
\end{array}
\end{equation}
We consider an initial sequence measuring $ Y_2$ and $ Y_5$. Measuring $ Y_2$:
\begin{equation}
\begin{array}{ll}
    S & =  \langle \pm Y_2, X_1 X_3 X_4, Z_3 Z_4 Z_5, Z_1 Y_3 Y_2 X_5  \rangle ,\\
    \bar{X} & =  X_4  X_5, \\ 
    \bar{Z} & = Z_1 Z_4.
\end{array}
\end{equation}
Followed by measuring $Y_5$:
 \begin{equation}
\begin{array}{ll}
    S & =  \langle \pm Y_2, X_1 X_3 X_4, \pm Y_5, Z_1 X_3 Z_4 Y_5  \rangle, \\
    \bar{X} & =  Z_3 Y_4, \\ 
    \bar{Z} & = Z_1 Z_4.
\end{array}
\end{equation}
Both logical operators remain well-defined.

Let us now consider the sequence $ Y_2,  Y_3, Y_4$. The latter corresponds to a diagonal of the code ($\bar{Y}$) and leads to the loss of the encoded information.\\
Measuring $Y_2$:
\begin{equation}
\begin{array}{ll}
    S & =  \langle \pm Y_2, X_1 X_3 X_4, Z_3 Z_4 Z_5, Z_1 Y_3 Y_2 X_5  \rangle, \\
    \bar{X} & =  X_4  X_5, \\ 
    \bar{Z} & = Z_1 Z_4.
\end{array}
\end{equation}
Followed by $Y_3$:
 \begin{equation}
\begin{array}{ll}
    S & =  \langle \pm Y_2, \pm Y_3, X_1 Y_3 Y_4 Z_5, Z_1 Y_2 Y_3 X_5 \rangle, \\
    \bar{X} & =  X_4 X_5, \\ 
    \bar{Z} & = Z_1 Z_4.
\end{array}
\end{equation}
Followed by $Y_4$:
 \begin{equation}
\begin{array}{ll}
    S & =  \langle \pm Y_2, \pm Y_3, X_1 Y_3 Y_4 Z_5, Z_1 Y_2 Y_3 X_5,  \rangle \\
    \bar{X} & =  \pm Y_4 \\ 
    \bar{Z} & = \pm Y_4.
\end{array}
\end{equation}
Both logical operators are replaced by $Y_4$. The logical subspace is no longer well-defined and the information is lost.

\subsection{Failure of logical $\bar{Y}$}
In this section, we check the validity of the threshold of $p_y=1$ for $\hat Y$-measurements by taking the limit $d\to \infty$ of Eq.~\eqref{eq: pfaily}. We start by rewriting this equation as the difference between two sums, i.e. 
\begin{equation}
\label{eq: pfaily2}
\begin{split}
        \mathcal{R}^{\rm{fail}}_{\bar{Y}}(n, p_y) =  & \sum_{k = 2d-1}^{n} 2 \binom{n-(2d-1)}{k - (2d-1)}  p_y^{k} (1- p_y)^{n-k}  \\
        & -  \sum_{k' = 4d-3}^{n} \binom{n-(4d-3)}{k' - (4d-3)}  p_y^{k'} (1- p_y)^{n-k'}. 
\end{split}
\end{equation}

Changing variables $\tilde{k} = k -(2d-1)$, $\tilde{n} = n -(2d-1)$, $\tilde{k'} = k -(4d-3)$, $\tilde{n'} = n' -(4d-3)$, we obtain a binomial distribution for each sum. Hence, in the limit of large $d$, both sums are well approximated by a normal distribution
\begin{equation}
\label{eq: pfaily3}
\begin{split}
        \mathcal{R}^{\rm{fail}}_{\bar{Y}}(n, p_y) \approx & p^{2d-1} \frac{e^{ -\frac{(\tilde{k} - \tilde{n}p)^{2}}{2\tilde{n}p (1 - p)}}}{ \sqrt{2\pi\tilde{n}p (1 - p)}} \\
        & - p^{4d-3} \frac{e^{- \frac{(\tilde{k'} - \tilde{n'}p)^{2}}{2\tilde{n'}p (1 - p)}}}{ \sqrt{2\pi\tilde{n'}p (1 - p)}}.
\end{split}
\end{equation}
In the limit  $d\to \infty$ this yields
\begin{equation}
    \lim_{d\to \infty }\mathcal{R}^{\rm{fail}}_{\bar{Y}}(n, p_y) = \left\lbrace 
    \begin{array}{cc}
        0  &  \forall \, \, 0\leq p < 1 \\
        1  &  \textrm{for} \quad p = 1
    \end{array}
    \right..
\end{equation}

\begin{figure}[hbt]
    \centering
    \includegraphics[width=1\columnwidth]{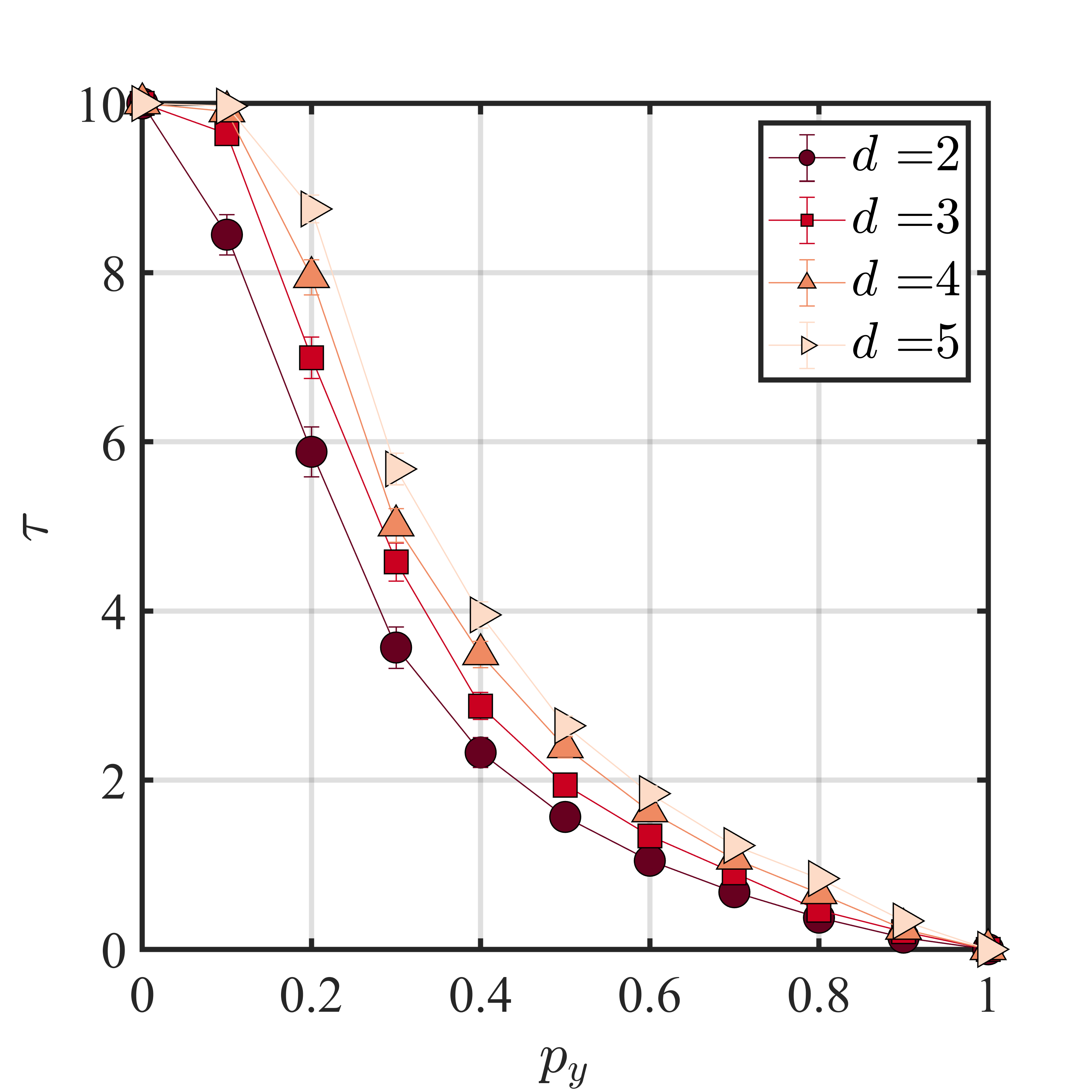}
    \caption{Lifetime of the logical information $\tau$ as a function of $p_y$ for different code distances $d$. The parameters underlying the simulation are $p_{\textrm{p}} = 0.1$, maximal time step $\Delta_{\rm{tmax}}$ = 11. We averaged over 1000 realizations. }
    \label{fig: app_life}
\end{figure}

\subsection{Dynamics: competition with stabilizer generator measurements}
\label{ap: threshold_y_dynamics}

Here, we briefly comment on the situation when we have a competition between measuring $\hat Y$ with probability $p_y$ and stabilizer $\hat P_l, \hat S_l$ with probability $p_s$. In Fig.~\ref{fig: app_life}, we show the lifetime of the logical information $\tau$ as a function of the probability $p_y$ for different code distances $d$, and a fixed low $p_s = 0.1$. We fixed the maximum runtime of the algorithm at $11$ rounds and averaged over 1000 realizations.
We see that the lines cross at $p_y =1 $ and that increasing the code distance increases the lifetime $\tau$. As a consequence, the threshold is given by $p_y = 1$. Note that this behavior is independent of the type of stabilizer generator considered as long as the joint probability $p_s$ is fixed.

\section{Validity of percolation description during the measurement-only dynamics}
\label{ap: validity_square}

\begin{figure*}[hbt]
    \centering
    \includegraphics[width=1\textwidth]{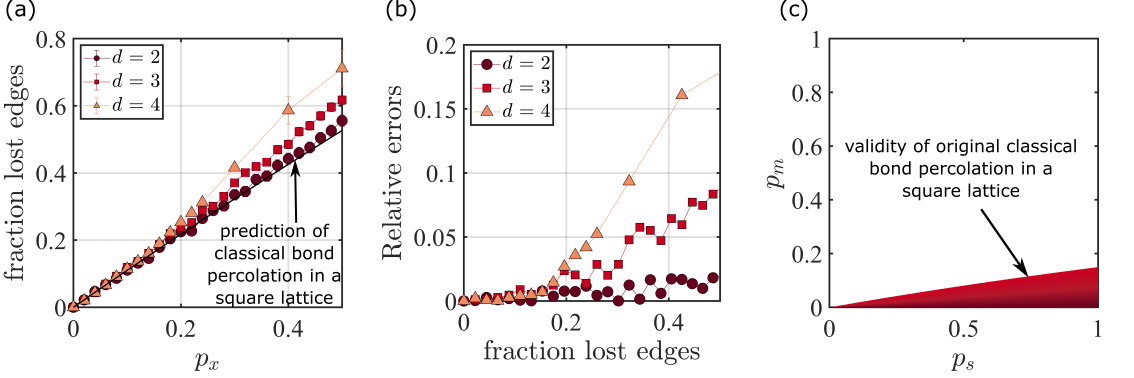}
    \caption{(a) Long time limit of the fraction of lost edges in the graph version of the planar surface code as a function of $p_x$ for a fix $p_{s} = 0.9$. The black line is the analytical prediction from Eq.~\eqref{eq: dynamic_fraction} with $\tilde{p}_{s} = p_{s}$. (b) Relative errors between the fraction of edges computed from the graph and the analytical prediction Eq.~\eqref{eq: dynamic_fraction} as a function of $p_x$. When the fraction exceeds 0.15, the relative errors grow quickly with the code distance. (c) Fraction of lost edges Eq.~\eqref{eq: dynamic_fraction} in the valid region of the phase diagram as a function of $p_x$ and $p_{s}$. We use the criteria obtained from (b), i.e., when the fraction is above 0.15 in the long-time limit, the bond percolation square lattice is not valid.}
    \label{fig: app_bond_perco_valid}
\end{figure*}

In this section, we focus more on the dynamics of lost edges (see main text). In particular, we wish to investigate the region of the phase diagram where the bond percolation on an original square lattice is valid. This region should correspond to a situation where the number of edges missing is kept low during the dynamics. Indeed, in this case, the number of missing edges per stabilizer can be approximated by one. Since a stabilizer generator measurement acting on a unique missing bond will restore it, it is clear that the bond percolation description on a square lattice becomes relevant. In order to check this approximation, we compute the long-time limit ratio of missing edges in the graph corresponding to the planar surface code.

In Fig.~\ref{fig: app_bond_perco_valid}(a), we show the fraction of missing edges as a function of $p_x$ for different code distances $d$ (see color code). The dots are numerical data averaged over 1000 realizations. The black line is the analytical prediction obtained from Eq.~\eqref{eq: dynamic_fraction}. We observe a discrepancy between the analytic and the numerical simulation when the fraction exceeds $\approx 0.15$. In Fig.~\ref{fig: app_bond_perco_valid}(b), we show the difference of the fraction between the simulation and the analytical prediction as a function of the fraction of lost edges. We confirm that when the fraction exceeds $\approx 0.15$ the discrepancy grows quickly with the system size. In Fig.~\ref{fig: app_bond_perco_valid}(c), we compute the region of the phase diagram where bond percolation on a 2D square lattice is valid. This region is far from the phase diagram boundary. Hence, we cannot use the usual percolation model to estimate the transition in our measurement-only dynamics. However, in order to address this challenge, we developed a generalized  percolation model based on the classical percolation theory to understand the dynamics of this system (see Sec.~\ref{sec: connection}).

\section{Critical exponent $\nu$ from a single round of $\hat X$-measurements}
\label{ap: critical_1_round}
\begin{figure}[hbt]
    \centering
    \includegraphics[width=1\columnwidth]{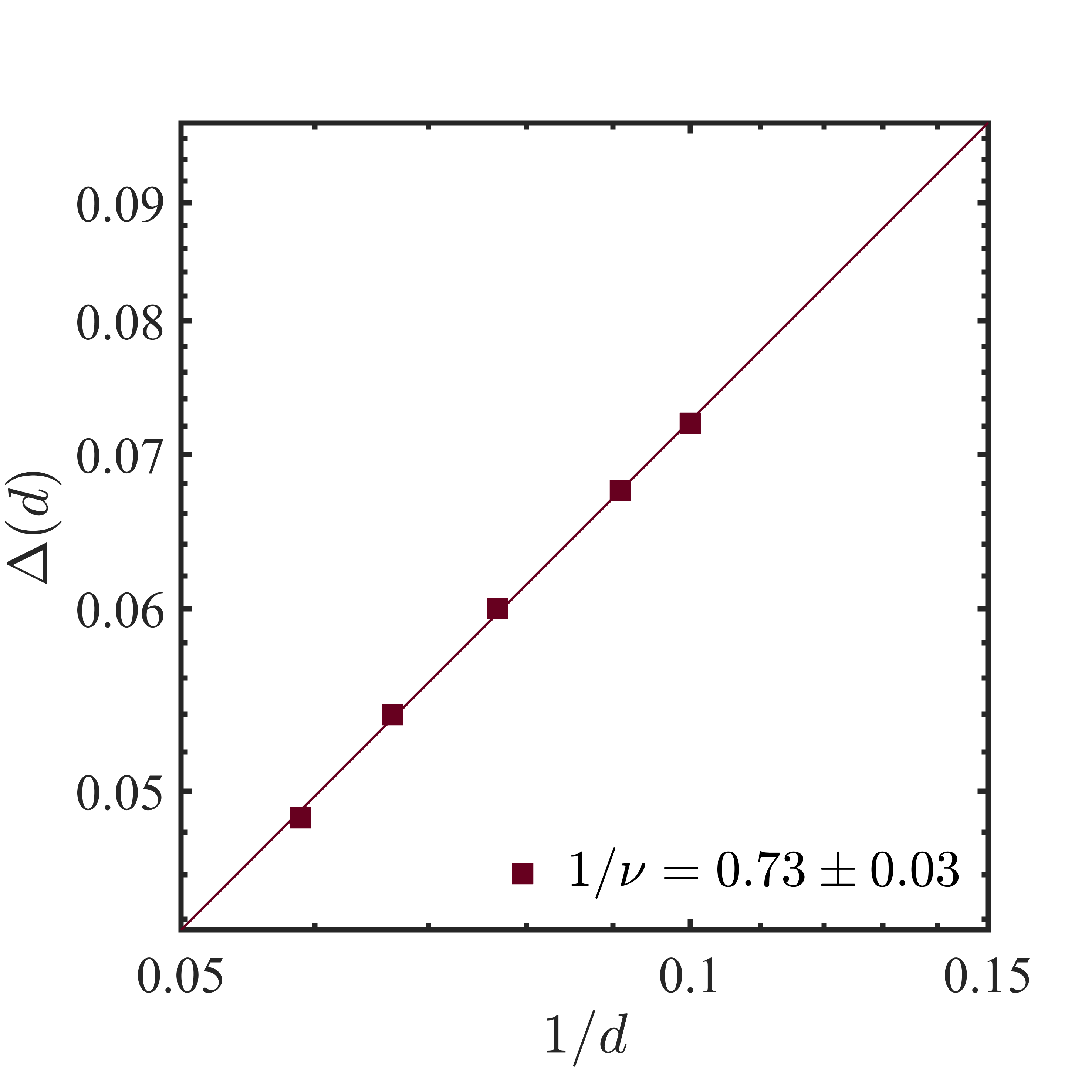}
    \caption{Critical behavior for a single round of $\hat X$-measurements. Standard deviation $\Delta(d)$ as a function of $1/d$. The line is a fit of the form $a/d^b$. Using Eq.~\eqref{eq: delta}, we identify $b = 1/\nu$ and obtain $1/\nu = 0.73 \pm 0.03$. }
    \label{fig: nu}
\end{figure}

Here, we examine the percolation behavior of the case of $\hat X$-measurements (see Sec.~\ref{sec: X_Z_meas_one_round}). In particular, we extract the critical exponent $\nu$ to corroborate the percolation transition. To do so, we use an algorithm developed in~\cite{DietrichStauffer2017Jan} to obtain the loss thresholds at finite code distance $d$ and in the thermodynamic limit $d \to \infty$.

We start by randomly drawing a set of  qubits to be measured in the  $\hat X$ basis with probability $p_1 = 1/2$. Then we check if the logical operators remain well-defined after the application of the local Pauli measurements. If this is the case, the process is repeated with a larger measurement probability rate $p_2 = p_1 + \zeta$. In case the logical information is lost, the measurement probability is reduced $p_2 = 1/2 - \zeta$. After each round, $\zeta \to \zeta + \zeta/2$, if the logical operators survive, otherwise $\zeta \to \zeta - \zeta/2$. The algorithm scales  logarithmically with the number of qubits in the lattice and stops when the number of measured qubits in two successive rounds does not change anymore. By repeating this operation, and then averaging, we obtain a distribution of the loss rates $p$ at which a logical operator is lost for the first time for a given fixed code distance $d$. The standard deviation of this distribution has been shown~\cite{DietrichStauffer2017Jan} to scale as 
\begin{equation}
\label{eq: delta}
    \Delta(d) \propto 1/d^{\frac{1}{\nu}}.
\end{equation}

In Fig.~\ref{fig: nu}, we show the standard deviation $\Delta(d)$ (square) as a function of $1/d$. The line is a fit of the form $a/d^{b}$. We identify $b = 1/\nu$ via Eq.~\eqref{eq: delta}, and we obtain $1/\nu = 0.73 \pm 0.03$. This value is in agreement with the analytical critical exponent for a 2D percolation transition with $\nu = \frac{4}{3}$.

\bibliography{main}

\begin{thebibliography}{72}%
\makeatletter
\providecommand \@ifxundefined [1]{%
 \@ifx{#1\undefined}
}%
\providecommand \@ifnum [1]{%
 \ifnum #1\expandafter \@firstoftwo
 \else \expandafter \@secondoftwo
 \fi
}%
\providecommand \@ifx [1]{%
 \ifx #1\expandafter \@firstoftwo
 \else \expandafter \@secondoftwo
 \fi
}%
\providecommand \natexlab [1]{#1}%
\providecommand \enquote  [1]{``#1''}%
\providecommand \bibnamefont  [1]{#1}%
\providecommand \bibfnamefont [1]{#1}%
\providecommand \citenamefont [1]{#1}%
\providecommand \href@noop [0]{\@secondoftwo}%
\providecommand \href [0]{\begingroup \@sanitize@url \@href}%
\providecommand \@href[1]{\@@startlink{#1}\@@href}%
\providecommand \@@href[1]{\endgroup#1\@@endlink}%
\providecommand \@sanitize@url [0]{\catcode `\\12\catcode `\$12\catcode
  `\&12\catcode `\#12\catcode `\^12\catcode `\_12\catcode `\%12\relax}%
\providecommand \@@startlink[1]{}%
\providecommand \@@endlink[0]{}%
\providecommand \url  [0]{\begingroup\@sanitize@url \@url }%
\providecommand \@url [1]{\endgroup\@href {#1}{\urlprefix }}%
\providecommand \urlprefix  [0]{URL }%
\providecommand \Eprint [0]{\href }%
\providecommand \doibase [0]{https://doi.org/}%
\providecommand \selectlanguage [0]{\@gobble}%
\providecommand \bibinfo  [0]{\@secondoftwo}%
\providecommand \bibfield  [0]{\@secondoftwo}%
\providecommand \translation [1]{[#1]}%
\providecommand \BibitemOpen [0]{}%
\providecommand \bibitemStop [0]{}%
\providecommand \bibitemNoStop [0]{.\EOS\space}%
\providecommand \EOS [0]{\spacefactor3000\relax}%
\providecommand \BibitemShut  [1]{\csname bibitem#1\endcsname}%
\let\auto@bib@innerbib\@empty
%</preamble>
\bibitem [{\citenamefont {Nielsen}\ and\ \citenamefont
  {Chuang}(2010)}]{Nielsen2010Dec}%
  \BibitemOpen
  \bibfield  {author} {\bibinfo {author} {\bibfnamefont {M.~A.}\ \bibnamefont
  {Nielsen}}\ and\ \bibinfo {author} {\bibfnamefont {I.~L.}\ \bibnamefont
  {Chuang}},\ }\href {https://doi.org/10.1017/CBO9780511976667} {\emph
  {\bibinfo {title} {{Quantum Computation and Quantum Information: 10th
  Anniversary Edition}}}}\ (\bibinfo  {publisher} {Cambridge University
  Press},\ \bibinfo {year} {2010})\BibitemShut {NoStop}%
\bibitem [{\citenamefont {Terhal}(2015)}]{Terhal2015Apr}%
  \BibitemOpen
  \bibfield  {author} {\bibinfo {author} {\bibfnamefont {B.~M.}\ \bibnamefont
  {Terhal}},\ }\href {https://doi.org/10.1103/RevModPhys.87.307} {\bibfield
  {journal} {\bibinfo  {journal} {Rev. Mod. Phys.}\ }\textbf {\bibinfo {volume}
  {87}},\ \bibinfo {pages} {307} (\bibinfo {year} {2015})}\BibitemShut
  {NoStop}%
\bibitem [{\citenamefont {Li}\ \emph {et~al.}(2018)\citenamefont {Li},
  \citenamefont {Chen},\ and\ \citenamefont {Fisher}}]{Li2018}%
  \BibitemOpen
  \bibfield  {author} {\bibinfo {author} {\bibfnamefont {Y.}~\bibnamefont
  {Li}}, \bibinfo {author} {\bibfnamefont {X.}~\bibnamefont {Chen}},\ and\
  \bibinfo {author} {\bibfnamefont {M.~P.~A.}\ \bibnamefont {Fisher}},\ }\href
  {https://doi.org/10.1103/PhysRevB.98.205136} {\bibfield  {journal} {\bibinfo
  {journal} {Phys. Rev. B}\ }\textbf {\bibinfo {volume} {98}},\ \bibinfo
  {pages} {205136} (\bibinfo {year} {2018})}\BibitemShut {NoStop}%
\bibitem [{\citenamefont {Chan}\ \emph {et~al.}(2019)\citenamefont {Chan},
  \citenamefont {Nandkishore}, \citenamefont {Pretko},\ and\ \citenamefont
  {Smith}}]{Chan2019}%
  \BibitemOpen
  \bibfield  {author} {\bibinfo {author} {\bibfnamefont {A.}~\bibnamefont
  {Chan}}, \bibinfo {author} {\bibfnamefont {R.~M.}\ \bibnamefont
  {Nandkishore}}, \bibinfo {author} {\bibfnamefont {M.}~\bibnamefont
  {Pretko}},\ and\ \bibinfo {author} {\bibfnamefont {G.}~\bibnamefont
  {Smith}},\ }\href {https://doi.org/10.1103/PhysRevB.99.224307} {\bibfield
  {journal} {\bibinfo  {journal} {Phys. Rev. B}\ }\textbf {\bibinfo {volume}
  {99}},\ \bibinfo {pages} {224307} (\bibinfo {year} {2019})}\BibitemShut
  {NoStop}%
\bibitem [{\citenamefont {Skinner}\ \emph {et~al.}(2019)\citenamefont
  {Skinner}, \citenamefont {Ruhman},\ and\ \citenamefont
  {Nahum}}]{Skinner2019}%
  \BibitemOpen
  \bibfield  {author} {\bibinfo {author} {\bibfnamefont {B.}~\bibnamefont
  {Skinner}}, \bibinfo {author} {\bibfnamefont {J.}~\bibnamefont {Ruhman}},\
  and\ \bibinfo {author} {\bibfnamefont {A.}~\bibnamefont {Nahum}},\ }\href
  {https://doi.org/10.1103/PhysRevX.9.031009} {\bibfield  {journal} {\bibinfo
  {journal} {Phys. Rev. X}\ }\textbf {\bibinfo {volume} {9}},\ \bibinfo {pages}
  {031009} (\bibinfo {year} {2019})}\BibitemShut {NoStop}%
\bibitem [{\citenamefont {Li}\ \emph {et~al.}(2019)\citenamefont {Li},
  \citenamefont {Chen},\ and\ \citenamefont {Fisher}}]{Li2019}%
  \BibitemOpen
  \bibfield  {author} {\bibinfo {author} {\bibfnamefont {Y.}~\bibnamefont
  {Li}}, \bibinfo {author} {\bibfnamefont {X.}~\bibnamefont {Chen}},\ and\
  \bibinfo {author} {\bibfnamefont {M.~P.~A.}\ \bibnamefont {Fisher}},\ }\href
  {https://doi.org/10.1103/PhysRevB.100.134306} {\bibfield  {journal} {\bibinfo
   {journal} {Phys. Rev. B}\ }\textbf {\bibinfo {volume} {100}},\ \bibinfo
  {pages} {134306} (\bibinfo {year} {2019})}\BibitemShut {NoStop}%
\bibitem [{\citenamefont {Jian}\ \emph {et~al.}(2020)\citenamefont {Jian},
  \citenamefont {You}, \citenamefont {Vasseur},\ and\ \citenamefont
  {Ludwig}}]{Jian2020}%
  \BibitemOpen
  \bibfield  {author} {\bibinfo {author} {\bibfnamefont {C.-M.}\ \bibnamefont
  {Jian}}, \bibinfo {author} {\bibfnamefont {Y.-Z.}\ \bibnamefont {You}},
  \bibinfo {author} {\bibfnamefont {R.}~\bibnamefont {Vasseur}},\ and\ \bibinfo
  {author} {\bibfnamefont {A.~W.~W.}\ \bibnamefont {Ludwig}},\ }\href
  {https://doi.org/10.1103/PhysRevB.101.104302} {\bibfield  {journal} {\bibinfo
   {journal} {Phys. Rev. B}\ }\textbf {\bibinfo {volume} {101}},\ \bibinfo
  {pages} {104302} (\bibinfo {year} {2020})}\BibitemShut {NoStop}%
\bibitem [{\citenamefont {Bao}\ \emph {et~al.}(2020)\citenamefont {Bao},
  \citenamefont {Choi},\ and\ \citenamefont {Altman}}]{Bao2020}%
  \BibitemOpen
  \bibfield  {author} {\bibinfo {author} {\bibfnamefont {Y.}~\bibnamefont
  {Bao}}, \bibinfo {author} {\bibfnamefont {S.}~\bibnamefont {Choi}},\ and\
  \bibinfo {author} {\bibfnamefont {E.}~\bibnamefont {Altman}},\ }\href
  {https://doi.org/10.1103/PhysRevB.101.104301} {\bibfield  {journal} {\bibinfo
   {journal} {Phys. Rev. B}\ }\textbf {\bibinfo {volume} {101}},\ \bibinfo
  {pages} {104301} (\bibinfo {year} {2020})}\BibitemShut {NoStop}%
\bibitem [{\citenamefont {Szyniszewski}\ \emph {et~al.}(2019)\citenamefont
  {Szyniszewski}, \citenamefont {Romito},\ and\ \citenamefont
  {Schomerus}}]{Szyniszewski2019}%
  \BibitemOpen
  \bibfield  {author} {\bibinfo {author} {\bibfnamefont {M.}~\bibnamefont
  {Szyniszewski}}, \bibinfo {author} {\bibfnamefont {A.}~\bibnamefont
  {Romito}},\ and\ \bibinfo {author} {\bibfnamefont {H.}~\bibnamefont
  {Schomerus}},\ }\href {https://doi.org/10.1103/PhysRevB.100.064204}
  {\bibfield  {journal} {\bibinfo  {journal} {Phys. Rev. B}\ }\textbf {\bibinfo
  {volume} {100}},\ \bibinfo {pages} {064204} (\bibinfo {year}
  {2019})}\BibitemShut {NoStop}%
\bibitem [{\citenamefont {Zabalo}\ \emph {et~al.}(2020)\citenamefont {Zabalo},
  \citenamefont {Gullans}, \citenamefont {Wilson}, \citenamefont
  {Gopalakrishnan}, \citenamefont {Huse},\ and\ \citenamefont
  {Pixley}}]{Zabalo2020}%
  \BibitemOpen
  \bibfield  {author} {\bibinfo {author} {\bibfnamefont {A.}~\bibnamefont
  {Zabalo}}, \bibinfo {author} {\bibfnamefont {M.~J.}\ \bibnamefont {Gullans}},
  \bibinfo {author} {\bibfnamefont {J.~H.}\ \bibnamefont {Wilson}}, \bibinfo
  {author} {\bibfnamefont {S.}~\bibnamefont {Gopalakrishnan}}, \bibinfo
  {author} {\bibfnamefont {D.~A.}\ \bibnamefont {Huse}},\ and\ \bibinfo
  {author} {\bibfnamefont {J.~H.}\ \bibnamefont {Pixley}},\ }\href
  {https://doi.org/10.1103/PhysRevB.101.060301} {\bibfield  {journal} {\bibinfo
   {journal} {Phys. Rev. B}\ }\textbf {\bibinfo {volume} {101}},\ \bibinfo
  {pages} {060301} (\bibinfo {year} {2020})}\BibitemShut {NoStop}%
\bibitem [{\citenamefont {Gullans}\ and\ \citenamefont
  {Huse}(2020{\natexlab{a}})}]{Gullans2020Aug}%
  \BibitemOpen
  \bibfield  {author} {\bibinfo {author} {\bibfnamefont {M.~J.}\ \bibnamefont
  {Gullans}}\ and\ \bibinfo {author} {\bibfnamefont {D.~A.}\ \bibnamefont
  {Huse}},\ }\href {https://doi.org/10.1103/PhysRevLett.125.070606} {\bibfield
  {journal} {\bibinfo  {journal} {Phys. Rev. Lett.}\ }\textbf {\bibinfo
  {volume} {125}},\ \bibinfo {pages} {070606} (\bibinfo {year}
  {2020}{\natexlab{a}})}\BibitemShut {NoStop}%
\bibitem [{\citenamefont {Gullans}\ and\ \citenamefont
  {Huse}(2020{\natexlab{b}})}]{Gullans2020Oct}%
  \BibitemOpen
  \bibfield  {author} {\bibinfo {author} {\bibfnamefont {M.~J.}\ \bibnamefont
  {Gullans}}\ and\ \bibinfo {author} {\bibfnamefont {D.~A.}\ \bibnamefont
  {Huse}},\ }\href {https://doi.org/10.1103/PhysRevX.10.041020} {\bibfield
  {journal} {\bibinfo  {journal} {Phys. Rev. X}\ }\textbf {\bibinfo {volume}
  {10}},\ \bibinfo {pages} {041020} (\bibinfo {year}
  {2020}{\natexlab{b}})}\BibitemShut {NoStop}%
\bibitem [{\citenamefont {Choi}\ \emph {et~al.}(2020)\citenamefont {Choi},
  \citenamefont {Bao}, \citenamefont {Qi},\ and\ \citenamefont
  {Altman}}]{Choi2020}%
  \BibitemOpen
  \bibfield  {author} {\bibinfo {author} {\bibfnamefont {S.}~\bibnamefont
  {Choi}}, \bibinfo {author} {\bibfnamefont {Y.}~\bibnamefont {Bao}}, \bibinfo
  {author} {\bibfnamefont {X.-L.}\ \bibnamefont {Qi}},\ and\ \bibinfo {author}
  {\bibfnamefont {E.}~\bibnamefont {Altman}},\ }\href
  {https://doi.org/10.1103/PhysRevLett.125.030505} {\bibfield  {journal}
  {\bibinfo  {journal} {Phys. Rev. Lett.}\ }\textbf {\bibinfo {volume} {125}},\
  \bibinfo {pages} {030505} (\bibinfo {year} {2020})}\BibitemShut {NoStop}%
\bibitem [{\citenamefont {Nahum}\ \emph {et~al.}(2021)\citenamefont {Nahum},
  \citenamefont {Roy}, \citenamefont {Skinner},\ and\ \citenamefont
  {Ruhman}}]{Nahum2021}%
  \BibitemOpen
  \bibfield  {author} {\bibinfo {author} {\bibfnamefont {A.}~\bibnamefont
  {Nahum}}, \bibinfo {author} {\bibfnamefont {S.}~\bibnamefont {Roy}}, \bibinfo
  {author} {\bibfnamefont {B.}~\bibnamefont {Skinner}},\ and\ \bibinfo {author}
  {\bibfnamefont {J.}~\bibnamefont {Ruhman}},\ }\href
  {https://doi.org/10.1103/PRXQuantum.2.010352} {\bibfield  {journal} {\bibinfo
   {journal} {PRX Quantum}\ }\textbf {\bibinfo {volume} {2}},\ \bibinfo {pages}
  {010352} (\bibinfo {year} {2021})}\BibitemShut {NoStop}%
\bibitem [{\citenamefont {Sang}\ and\ \citenamefont
  {Hsieh}(2021)}]{Sang2020Apr}%
  \BibitemOpen
  \bibfield  {author} {\bibinfo {author} {\bibfnamefont {S.}~\bibnamefont
  {Sang}}\ and\ \bibinfo {author} {\bibfnamefont {T.~H.}\ \bibnamefont
  {Hsieh}},\ }\href {https://doi.org/10.1103/PhysRevResearch.3.023200}
  {\bibfield  {journal} {\bibinfo  {journal} {Phys. Rev. Res.}\ }\textbf
  {\bibinfo {volume} {3}},\ \bibinfo {pages} {023200} (\bibinfo {year}
  {2021})}\BibitemShut {NoStop}%
\bibitem [{\citenamefont {Bao}\ \emph {et~al.}(2021)\citenamefont {Bao},
  \citenamefont {Choi},\ and\ \citenamefont {Altman}}]{Bao2021Feb}%
  \BibitemOpen
  \bibfield  {author} {\bibinfo {author} {\bibfnamefont {Y.}~\bibnamefont
  {Bao}}, \bibinfo {author} {\bibfnamefont {S.}~\bibnamefont {Choi}},\ and\
  \bibinfo {author} {\bibfnamefont {E.}~\bibnamefont {Altman}},\ }\href
  {https://arxiv.org/abs/2102.09164v1} {\bibfield  {journal} {\bibinfo
  {journal} {arXiv}\ } (\bibinfo {year} {2021})},\ \Eprint
  {https://arxiv.org/abs/2102.09164} {2102.09164} \BibitemShut {NoStop}%
\bibitem [{\citenamefont {Lavasani}\ \emph
  {et~al.}(2021{\natexlab{a}})\citenamefont {Lavasani}, \citenamefont
  {Alavirad},\ and\ \citenamefont {Barkeshli}}]{Lavasani2021Mar}%
  \BibitemOpen
  \bibfield  {author} {\bibinfo {author} {\bibfnamefont {A.}~\bibnamefont
  {Lavasani}}, \bibinfo {author} {\bibfnamefont {Y.}~\bibnamefont {Alavirad}},\
  and\ \bibinfo {author} {\bibfnamefont {M.}~\bibnamefont {Barkeshli}},\ }\href
  {https://doi.org/10.1038/s41567-020-01112-z} {\bibfield  {journal} {\bibinfo
  {journal} {Nat. Phys.}\ }\textbf {\bibinfo {volume} {17}},\ \bibinfo {pages}
  {342} (\bibinfo {year} {2021}{\natexlab{a}})}\BibitemShut {NoStop}%
\bibitem [{\citenamefont {Sierant}\ \emph {et~al.}(2022)\citenamefont
  {Sierant}, \citenamefont {Chiriac{\ifmmode\grave{o}\else\`{o}\fi}},
  \citenamefont {Surace}, \citenamefont {Sharma}, \citenamefont {Turkeshi},
  \citenamefont {Dalmonte}, \citenamefont {Fazio},\ and\ \citenamefont
  {Pagano}}]{Sierant2022Feb}%
  \BibitemOpen
  \bibfield  {author} {\bibinfo {author} {\bibfnamefont {P.}~\bibnamefont
  {Sierant}}, \bibinfo {author} {\bibfnamefont {G.}~\bibnamefont
  {Chiriac{\ifmmode\grave{o}\else\`{o}\fi}}}, \bibinfo {author} {\bibfnamefont
  {F.~M.}\ \bibnamefont {Surace}}, \bibinfo {author} {\bibfnamefont
  {S.}~\bibnamefont {Sharma}}, \bibinfo {author} {\bibfnamefont
  {X.}~\bibnamefont {Turkeshi}}, \bibinfo {author} {\bibfnamefont
  {M.}~\bibnamefont {Dalmonte}}, \bibinfo {author} {\bibfnamefont
  {R.}~\bibnamefont {Fazio}},\ and\ \bibinfo {author} {\bibfnamefont
  {G.}~\bibnamefont {Pagano}},\ }\href
  {https://doi.org/10.22331/q-2022-02-02-638} {\bibfield  {journal} {\bibinfo
  {journal} {Quantum}\ }\textbf {\bibinfo {volume} {6}},\ \bibinfo {pages}
  {638} (\bibinfo {year} {2022})}\BibitemShut {NoStop}%
\bibitem [{\citenamefont {{Zhu}}\ \emph {et~al.}(2022)\citenamefont {{Zhu}},
  \citenamefont {{Tantivasadakarn}}, \citenamefont {{Vishwanath}},
  \citenamefont {{Trebst}},\ and\ \citenamefont {{Verresen}}}]{Guoyi2022}%
  \BibitemOpen
  \bibfield  {author} {\bibinfo {author} {\bibfnamefont {G.-Y.}\ \bibnamefont
  {{Zhu}}}, \bibinfo {author} {\bibfnamefont {N.}~\bibnamefont
  {{Tantivasadakarn}}}, \bibinfo {author} {\bibfnamefont {A.}~\bibnamefont
  {{Vishwanath}}}, \bibinfo {author} {\bibfnamefont {S.}~\bibnamefont
  {{Trebst}}},\ and\ \bibinfo {author} {\bibfnamefont {R.}~\bibnamefont
  {{Verresen}}},\ }\href {https://doi.org/10.48550/arXiv.2208.11136} {}
  (\bibinfo {year} {2022}),\ \Eprint {https://arxiv.org/abs/2208.11136}
  {arXiv:2208.11136} \BibitemShut {NoStop}%
\bibitem [{\citenamefont {Fan}\ \emph {et~al.}(2020)\citenamefont {Fan},
  \citenamefont {Vijay}, \citenamefont {Vishwanath},\ and\ \citenamefont
  {You}}]{Fan2020}%
  \BibitemOpen
  \bibfield  {author} {\bibinfo {author} {\bibfnamefont {R.}~\bibnamefont
  {Fan}}, \bibinfo {author} {\bibfnamefont {S.}~\bibnamefont {Vijay}}, \bibinfo
  {author} {\bibfnamefont {A.}~\bibnamefont {Vishwanath}},\ and\ \bibinfo
  {author} {\bibfnamefont {Y.-Z.}\ \bibnamefont {You}},\ }\href
  {https://arxiv.org/abs/2002.12385v1} {\bibfield  {journal} {\bibinfo
  {journal} {arXiv}\ } (\bibinfo {year} {2020})},\ \Eprint
  {https://arxiv.org/abs/2002.12385} {2002.12385} \BibitemShut {NoStop}%
\bibitem [{\citenamefont {Fisher}\ \emph {et~al.}(2023)\citenamefont {Fisher},
  \citenamefont {Khemani}, \citenamefont {Nahum},\ and\ \citenamefont
  {Vijay}}]{circuitreview}%
  \BibitemOpen
  \bibfield  {author} {\bibinfo {author} {\bibfnamefont {M.~P.}\ \bibnamefont
  {Fisher}}, \bibinfo {author} {\bibfnamefont {V.}~\bibnamefont {Khemani}},
  \bibinfo {author} {\bibfnamefont {A.}~\bibnamefont {Nahum}},\ and\ \bibinfo
  {author} {\bibfnamefont {S.}~\bibnamefont {Vijay}},\ }\href
  {https://doi.org/10.1146/annurev-conmatphys-031720-030658} {\bibfield
  {journal} {\bibinfo  {journal} {Annual Review of Condensed Matter Physics}\
  }\textbf {\bibinfo {volume} {14}},\ \bibinfo {pages} {335} (\bibinfo {year}
  {2023})},\ \Eprint
  {https://arxiv.org/abs/https://doi.org/10.1146/annurev-conmatphys-031720-030658}
  {https://doi.org/10.1146/annurev-conmatphys-031720-030658} \BibitemShut
  {NoStop}%
\bibitem [{\citenamefont {Szyniszewski}\ \emph {et~al.}(2020)\citenamefont
  {Szyniszewski}, \citenamefont {Romito},\ and\ \citenamefont
  {Schomerus}}]{Romito2020}%
  \BibitemOpen
  \bibfield  {author} {\bibinfo {author} {\bibfnamefont {M.}~\bibnamefont
  {Szyniszewski}}, \bibinfo {author} {\bibfnamefont {A.}~\bibnamefont
  {Romito}},\ and\ \bibinfo {author} {\bibfnamefont {H.}~\bibnamefont
  {Schomerus}},\ }\href {https://doi.org/10.1103/PhysRevLett.125.210602}
  {\bibfield  {journal} {\bibinfo  {journal} {Phys. Rev. Lett.}\ }\textbf
  {\bibinfo {volume} {125}},\ \bibinfo {pages} {210602} (\bibinfo {year}
  {2020})}\BibitemShut {NoStop}%
\bibitem [{\citenamefont {{Morral-Yepes}}\ \emph {et~al.}(2023)\citenamefont
  {{Morral-Yepes}}, \citenamefont {{Pollmann}},\ and\ \citenamefont
  {{Lovas}}}]{Morral2022}%
  \BibitemOpen
  \bibfield  {author} {\bibinfo {author} {\bibfnamefont {R.}~\bibnamefont
  {{Morral-Yepes}}}, \bibinfo {author} {\bibfnamefont {F.}~\bibnamefont
  {{Pollmann}}},\ and\ \bibinfo {author} {\bibfnamefont {I.}~\bibnamefont
  {{Lovas}}},\ }\bibfield  {journal} {\bibinfo  {journal} {arXiv e-prints}\
  }\href {https://doi.org/10.48550/arXiv.2302.14551}
  {10.48550/arXiv.2302.14551} (\bibinfo {year} {2023})\BibitemShut {NoStop}%
\bibitem [{\citenamefont {Turkeshi}(2022)}]{turkeshi2022measurement}%
  \BibitemOpen
  \bibfield  {author} {\bibinfo {author} {\bibfnamefont {X.}~\bibnamefont
  {Turkeshi}},\ }\href {https://doi.org/10.1103/PhysRevB.106.144313} {\bibfield
   {journal} {\bibinfo  {journal} {Phys. Rev. B}\ }\textbf {\bibinfo {volume}
  {106}},\ \bibinfo {pages} {144313} (\bibinfo {year} {2022})}\BibitemShut
  {NoStop}%
\bibitem [{\citenamefont {Zabalo}\ \emph {et~al.}(2022)\citenamefont {Zabalo},
  \citenamefont {Gullans}, \citenamefont {Wilson}, \citenamefont {Vasseur},
  \citenamefont {Ludwig}, \citenamefont {Gopalakrishnan}, \citenamefont
  {Huse},\ and\ \citenamefont {Pixley}}]{Zabalo2022}%
  \BibitemOpen
  \bibfield  {author} {\bibinfo {author} {\bibfnamefont {A.}~\bibnamefont
  {Zabalo}}, \bibinfo {author} {\bibfnamefont {M.~J.}\ \bibnamefont {Gullans}},
  \bibinfo {author} {\bibfnamefont {J.~H.}\ \bibnamefont {Wilson}}, \bibinfo
  {author} {\bibfnamefont {R.}~\bibnamefont {Vasseur}}, \bibinfo {author}
  {\bibfnamefont {A.~W.~W.}\ \bibnamefont {Ludwig}}, \bibinfo {author}
  {\bibfnamefont {S.}~\bibnamefont {Gopalakrishnan}}, \bibinfo {author}
  {\bibfnamefont {D.~A.}\ \bibnamefont {Huse}},\ and\ \bibinfo {author}
  {\bibfnamefont {J.~H.}\ \bibnamefont {Pixley}},\ }\href
  {https://doi.org/10.1103/PhysRevLett.128.050602} {\bibfield  {journal}
  {\bibinfo  {journal} {Phys. Rev. Lett.}\ }\textbf {\bibinfo {volume} {128}},\
  \bibinfo {pages} {050602} (\bibinfo {year} {2022})}\BibitemShut {NoStop}%
\bibitem [{\citenamefont {Cao}\ \emph {et~al.}(2019)\citenamefont {Cao},
  \citenamefont {Tilloy},\ and\ \citenamefont {De~Luca}}]{Cao2019}%
  \BibitemOpen
  \bibfield  {author} {\bibinfo {author} {\bibfnamefont {X.}~\bibnamefont
  {Cao}}, \bibinfo {author} {\bibfnamefont {A.}~\bibnamefont {Tilloy}},\ and\
  \bibinfo {author} {\bibfnamefont {A.}~\bibnamefont {De~Luca}},\ }\href
  {https://doi.org/10.21468/SciPostPhys.7.2.024} {\bibfield  {journal}
  {\bibinfo  {journal} {SciPost Phys.}\ }\textbf {\bibinfo {volume} {7}},\
  \bibinfo {pages} {024} (\bibinfo {year} {2019})}\BibitemShut {NoStop}%
\bibitem [{\citenamefont {Alberton}\ \emph {et~al.}(2021)\citenamefont
  {Alberton}, \citenamefont {Buchhold},\ and\ \citenamefont
  {Diehl}}]{Alberton2021}%
  \BibitemOpen
  \bibfield  {author} {\bibinfo {author} {\bibfnamefont {O.}~\bibnamefont
  {Alberton}}, \bibinfo {author} {\bibfnamefont {M.}~\bibnamefont {Buchhold}},\
  and\ \bibinfo {author} {\bibfnamefont {S.}~\bibnamefont {Diehl}},\ }\href
  {https://doi.org/10.1103/PhysRevLett.126.170602} {\bibfield  {journal}
  {\bibinfo  {journal} {Phys. Rev. Lett.}\ }\textbf {\bibinfo {volume} {126}},\
  \bibinfo {pages} {170602} (\bibinfo {year} {2021})}\BibitemShut {NoStop}%
\bibitem [{\citenamefont {Buchhold}\ \emph {et~al.}(2021)\citenamefont
  {Buchhold}, \citenamefont {Minoguchi}, \citenamefont {Altland},\ and\
  \citenamefont {Diehl}}]{Buchhold2021}%
  \BibitemOpen
  \bibfield  {author} {\bibinfo {author} {\bibfnamefont {M.}~\bibnamefont
  {Buchhold}}, \bibinfo {author} {\bibfnamefont {Y.}~\bibnamefont {Minoguchi}},
  \bibinfo {author} {\bibfnamefont {A.}~\bibnamefont {Altland}},\ and\ \bibinfo
  {author} {\bibfnamefont {S.}~\bibnamefont {Diehl}},\ }\href
  {https://arxiv.org/abs/2102.08381v2} {\bibfield  {journal} {\bibinfo
  {journal} {arXiv}\ } (\bibinfo {year} {2021})},\ \Eprint
  {https://arxiv.org/abs/2102.08381} {2102.08381} \BibitemShut {NoStop}%
\bibitem [{\citenamefont {Botzung}\ \emph {et~al.}(2021)\citenamefont
  {Botzung}, \citenamefont {Diehl},\ and\ \citenamefont
  {M{\ifmmode\ddot{u}\else\"{u}\fi}ller}}]{Botzung2021Nov}%
  \BibitemOpen
  \bibfield  {author} {\bibinfo {author} {\bibfnamefont {T.}~\bibnamefont
  {Botzung}}, \bibinfo {author} {\bibfnamefont {S.}~\bibnamefont {Diehl}},\
  and\ \bibinfo {author} {\bibfnamefont {M.}~\bibnamefont
  {M{\ifmmode\ddot{u}\else\"{u}\fi}ller}},\ }\href
  {https://doi.org/10.1103/PhysRevB.104.184422} {\bibfield  {journal} {\bibinfo
   {journal} {Phys. Rev. B}\ }\textbf {\bibinfo {volume} {104}},\ \bibinfo
  {pages} {184422} (\bibinfo {year} {2021})}\BibitemShut {NoStop}%
\bibitem [{\citenamefont {Piccitto}\ \emph {et~al.}(2022)\citenamefont
  {Piccitto}, \citenamefont {Russomanno},\ and\ \citenamefont
  {Rossini}}]{Piccitto2022Feb}%
  \BibitemOpen
  \bibfield  {author} {\bibinfo {author} {\bibfnamefont {G.}~\bibnamefont
  {Piccitto}}, \bibinfo {author} {\bibfnamefont {A.}~\bibnamefont
  {Russomanno}},\ and\ \bibinfo {author} {\bibfnamefont {D.}~\bibnamefont
  {Rossini}},\ }\href {https://doi.org/10.1103/PhysRevB.105.064305} {\bibfield
  {journal} {\bibinfo  {journal} {Phys. Rev. B}\ }\textbf {\bibinfo {volume}
  {105}},\ \bibinfo {pages} {064305} (\bibinfo {year} {2022})}\BibitemShut
  {NoStop}%
\bibitem [{\citenamefont {Fuji}\ and\ \citenamefont
  {Ashida}(2020)}]{Fuji2020Aug}%
  \BibitemOpen
  \bibfield  {author} {\bibinfo {author} {\bibfnamefont {Y.}~\bibnamefont
  {Fuji}}\ and\ \bibinfo {author} {\bibfnamefont {Y.}~\bibnamefont {Ashida}},\
  }\href {https://doi.org/10.1103/PhysRevB.102.054302} {\bibfield  {journal}
  {\bibinfo  {journal} {Phys. Rev. B}\ }\textbf {\bibinfo {volume} {102}},\
  \bibinfo {pages} {054302} (\bibinfo {year} {2020})}\BibitemShut {NoStop}%
\bibitem [{\citenamefont {Fuji}\ and\ \citenamefont
  {Ashida}(2021)}]{Fuji2021Feb}%
  \BibitemOpen
  \bibfield  {author} {\bibinfo {author} {\bibfnamefont {Y.}~\bibnamefont
  {Fuji}}\ and\ \bibinfo {author} {\bibfnamefont {Y.}~\bibnamefont {Ashida}},\
  }\href {https://doi.org/10.1103/PhysRevB.103.059901} {\bibfield  {journal}
  {\bibinfo  {journal} {Phys. Rev. B}\ }\textbf {\bibinfo {volume} {103}},\
  \bibinfo {pages} {059901} (\bibinfo {year} {2021})}\BibitemShut {NoStop}%
\bibitem [{\citenamefont {M{\ifmmode\ddot{u}\else\"{u}\fi}ller}\ \emph
  {et~al.}(2022)\citenamefont {M{\ifmmode\ddot{u}\else\"{u}\fi}ller},
  \citenamefont {Diehl},\ and\ \citenamefont {Buchhold}}]{Muller2022Jan}%
  \BibitemOpen
  \bibfield  {author} {\bibinfo {author} {\bibfnamefont {T.}~\bibnamefont
  {M{\ifmmode\ddot{u}\else\"{u}\fi}ller}}, \bibinfo {author} {\bibfnamefont
  {S.}~\bibnamefont {Diehl}},\ and\ \bibinfo {author} {\bibfnamefont
  {M.}~\bibnamefont {Buchhold}},\ }\href
  {https://doi.org/10.1103/PhysRevLett.128.010605} {\bibfield  {journal}
  {\bibinfo  {journal} {Phys. Rev. Lett.}\ }\textbf {\bibinfo {volume} {128}},\
  \bibinfo {pages} {010605} (\bibinfo {year} {2022})}\BibitemShut {NoStop}%
\bibitem [{\citenamefont {Block}\ \emph {et~al.}(2022)\citenamefont {Block},
  \citenamefont {Bao}, \citenamefont {Choi}, \citenamefont {Altman},\ and\
  \citenamefont {Yao}}]{Block2022Jan}%
  \BibitemOpen
  \bibfield  {author} {\bibinfo {author} {\bibfnamefont {M.}~\bibnamefont
  {Block}}, \bibinfo {author} {\bibfnamefont {Y.}~\bibnamefont {Bao}}, \bibinfo
  {author} {\bibfnamefont {S.}~\bibnamefont {Choi}}, \bibinfo {author}
  {\bibfnamefont {E.}~\bibnamefont {Altman}},\ and\ \bibinfo {author}
  {\bibfnamefont {N.~Y.}\ \bibnamefont {Yao}},\ }\href
  {https://doi.org/10.1103/PhysRevLett.128.010604} {\bibfield  {journal}
  {\bibinfo  {journal} {Phys. Rev. Lett.}\ }\textbf {\bibinfo {volume} {128}},\
  \bibinfo {pages} {010604} (\bibinfo {year} {2022})}\BibitemShut {NoStop}%
\bibitem [{\citenamefont {Zhang}\ \emph {et~al.}(2022)\citenamefont {Zhang},
  \citenamefont {Liu}, \citenamefont {Jian},\ and\ \citenamefont
  {Chen}}]{Zhang_2022}%
  \BibitemOpen
  \bibfield  {author} {\bibinfo {author} {\bibfnamefont {P.}~\bibnamefont
  {Zhang}}, \bibinfo {author} {\bibfnamefont {C.}~\bibnamefont {Liu}}, \bibinfo
  {author} {\bibfnamefont {S.-K.}\ \bibnamefont {Jian}},\ and\ \bibinfo
  {author} {\bibfnamefont {X.}~\bibnamefont {Chen}},\ }\href
  {https://doi.org/10.22331/q-2022-05-27-723} {\bibfield  {journal} {\bibinfo
  {journal} {Quantum}\ }\textbf {\bibinfo {volume} {6}},\ \bibinfo {pages}
  {723} (\bibinfo {year} {2022})}\BibitemShut {NoStop}%
\bibitem [{\citenamefont {Turkeshi}\ \emph {et~al.}(2021)\citenamefont
  {Turkeshi}, \citenamefont {Biella}, \citenamefont {Fazio}, \citenamefont
  {Dalmonte},\ and\ \citenamefont {Schir\'o}}]{turkeshi2021measurementinduced}%
  \BibitemOpen
  \bibfield  {author} {\bibinfo {author} {\bibfnamefont {X.}~\bibnamefont
  {Turkeshi}}, \bibinfo {author} {\bibfnamefont {A.}~\bibnamefont {Biella}},
  \bibinfo {author} {\bibfnamefont {R.}~\bibnamefont {Fazio}}, \bibinfo
  {author} {\bibfnamefont {M.}~\bibnamefont {Dalmonte}},\ and\ \bibinfo
  {author} {\bibfnamefont {M.}~\bibnamefont {Schir\'o}},\ }\href
  {https://doi.org/10.1103/PhysRevB.103.224210} {\bibfield  {journal} {\bibinfo
   {journal} {Phys. Rev. B}\ }\textbf {\bibinfo {volume} {103}},\ \bibinfo
  {pages} {224210} (\bibinfo {year} {2021})}\BibitemShut {NoStop}%
\bibitem [{\citenamefont {{Poboiko}}\ \emph {et~al.}(2023)\citenamefont
  {{Poboiko}}, \citenamefont {{P{\"o}pperl}}, \citenamefont {{Gornyi}},\ and\
  \citenamefont {{Mirlin}}}]{poboiko}%
  \BibitemOpen
  \bibfield  {author} {\bibinfo {author} {\bibfnamefont {I.}~\bibnamefont
  {{Poboiko}}}, \bibinfo {author} {\bibfnamefont {P.}~\bibnamefont
  {{P{\"o}pperl}}}, \bibinfo {author} {\bibfnamefont {I.~V.}\ \bibnamefont
  {{Gornyi}}},\ and\ \bibinfo {author} {\bibfnamefont {A.~D.}\ \bibnamefont
  {{Mirlin}}},\ }\bibfield  {journal} {\bibinfo  {journal} {arXiv e-prints}\
  }\href {https://doi.org/10.48550/arXiv.2304.03138}
  {10.48550/arXiv.2304.03138} (\bibinfo {year} {2023})\BibitemShut {NoStop}%
\bibitem [{\citenamefont {Doggen}\ \emph {et~al.}(2021)\citenamefont {Doggen},
  \citenamefont {Gefen}, \citenamefont {Gornyi}, \citenamefont {Mirlin},\ and\
  \citenamefont {Polyakov}}]{Doggen2021}%
  \BibitemOpen
  \bibfield  {author} {\bibinfo {author} {\bibfnamefont {E.~V.~H.}\
  \bibnamefont {Doggen}}, \bibinfo {author} {\bibfnamefont {Y.}~\bibnamefont
  {Gefen}}, \bibinfo {author} {\bibfnamefont {I.~V.}\ \bibnamefont {Gornyi}},
  \bibinfo {author} {\bibfnamefont {A.~D.}\ \bibnamefont {Mirlin}},\ and\
  \bibinfo {author} {\bibfnamefont {D.~G.}\ \bibnamefont {Polyakov}},\ }\href
  {https://arxiv.org/abs/2104.10451v1} {\bibfield  {journal} {\bibinfo
  {journal} {arXiv}\ } (\bibinfo {year} {2021})},\ \Eprint
  {https://arxiv.org/abs/2104.10451} {2104.10451} \BibitemShut {NoStop}%
\bibitem [{\citenamefont {Poboiko}\ \emph {et~al.}(2023)\citenamefont
  {Poboiko}, \citenamefont {Gornyi},\ and\ \citenamefont
  {Mirlin}}]{poboiko2023measurementinduced}%
  \BibitemOpen
  \bibfield  {author} {\bibinfo {author} {\bibfnamefont {I.}~\bibnamefont
  {Poboiko}}, \bibinfo {author} {\bibfnamefont {I.~V.}\ \bibnamefont
  {Gornyi}},\ and\ \bibinfo {author} {\bibfnamefont {A.~D.}\ \bibnamefont
  {Mirlin}},\ }\href@noop {} {\bibinfo {title} {Measurement-induced phase
  transition for free fermions above one dimension}} (\bibinfo {year} {2023}),\
  \Eprint {https://arxiv.org/abs/2309.12405} {arXiv:2309.12405 [quant-ph]}
  \BibitemShut {NoStop}%
\bibitem [{\citenamefont {P\"opperl}\ \emph {et~al.}(2023)\citenamefont
  {P\"opperl}, \citenamefont {Gornyi},\ and\ \citenamefont {Gefen}}]{Popperl}%
  \BibitemOpen
  \bibfield  {author} {\bibinfo {author} {\bibfnamefont {P.}~\bibnamefont
  {P\"opperl}}, \bibinfo {author} {\bibfnamefont {I.~V.}\ \bibnamefont
  {Gornyi}},\ and\ \bibinfo {author} {\bibfnamefont {Y.}~\bibnamefont
  {Gefen}},\ }\href {https://doi.org/10.1103/PhysRevB.107.174203} {\bibfield
  {journal} {\bibinfo  {journal} {Phys. Rev. B}\ }\textbf {\bibinfo {volume}
  {107}},\ \bibinfo {pages} {174203} (\bibinfo {year} {2023})}\BibitemShut
  {NoStop}%
\bibitem [{\citenamefont {{Chahine}}\ and\ \citenamefont
  {{Buchhold}}(2023)}]{Chahine2023}%
  \BibitemOpen
  \bibfield  {author} {\bibinfo {author} {\bibfnamefont {K.}~\bibnamefont
  {{Chahine}}}\ and\ \bibinfo {author} {\bibfnamefont {M.}~\bibnamefont
  {{Buchhold}}},\ }\href {https://doi.org/10.48550/arXiv.2309.12391} {\bibfield
   {journal} {\bibinfo  {journal} {arXiv e-prints}\ ,\ \bibinfo {eid}
  {arXiv:2309.12391}} (\bibinfo {year} {2023})}\BibitemShut {NoStop}%
\bibitem [{\citenamefont {Tirrito}\ \emph {et~al.}(2023)\citenamefont
  {Tirrito}, \citenamefont {Santini}, \citenamefont {Fazio},\ and\
  \citenamefont {Collura}}]{tirrito2023}%
  \BibitemOpen
  \bibfield  {author} {\bibinfo {author} {\bibfnamefont {E.}~\bibnamefont
  {Tirrito}}, \bibinfo {author} {\bibfnamefont {A.}~\bibnamefont {Santini}},
  \bibinfo {author} {\bibfnamefont {R.}~\bibnamefont {Fazio}},\ and\ \bibinfo
  {author} {\bibfnamefont {M.}~\bibnamefont {Collura}},\ }\href
  {https://doi.org/10.21468/SciPostPhys.15.3.096} {\bibfield  {journal}
  {\bibinfo  {journal} {SciPost Phys.}\ }\textbf {\bibinfo {volume} {15}},\
  \bibinfo {pages} {096} (\bibinfo {year} {2023})}\BibitemShut {NoStop}%
\bibitem [{\citenamefont {Vovk}\ and\ \citenamefont
  {Pichler}(2022)}]{Vovk2022Jun}%
  \BibitemOpen
  \bibfield  {author} {\bibinfo {author} {\bibfnamefont {T.}~\bibnamefont
  {Vovk}}\ and\ \bibinfo {author} {\bibfnamefont {H.}~\bibnamefont {Pichler}},\
  }\href {https://doi.org/10.1103/PhysRevLett.128.243601} {\bibfield  {journal}
  {\bibinfo  {journal} {Phys. Rev. Lett.}\ }\textbf {\bibinfo {volume} {128}},\
  \bibinfo {pages} {243601} (\bibinfo {year} {2022})}\BibitemShut {NoStop}%
\bibitem [{\citenamefont {Minoguchi}\ \emph {et~al.}(2022)\citenamefont
  {Minoguchi}, \citenamefont {Rabl},\ and\ \citenamefont
  {Buchhold}}]{Minoguchi2021}%
  \BibitemOpen
  \bibfield  {author} {\bibinfo {author} {\bibfnamefont {Y.}~\bibnamefont
  {Minoguchi}}, \bibinfo {author} {\bibfnamefont {P.}~\bibnamefont {Rabl}},\
  and\ \bibinfo {author} {\bibfnamefont {M.}~\bibnamefont {Buchhold}},\ }\href
  {https://doi.org/10.21468/SciPostPhys.12.1.009} {\bibfield  {journal}
  {\bibinfo  {journal} {SciPost Phys.}\ }\textbf {\bibinfo {volume} {12}},\
  \bibinfo {pages} {009} (\bibinfo {year} {2022})}\BibitemShut {NoStop}%
\bibitem [{\citenamefont {Ladewig}\ \emph {et~al.}(2022)\citenamefont
  {Ladewig}, \citenamefont {Diehl},\ and\ \citenamefont {Buchhold}}]{Ladewig}%
  \BibitemOpen
  \bibfield  {author} {\bibinfo {author} {\bibfnamefont {B.}~\bibnamefont
  {Ladewig}}, \bibinfo {author} {\bibfnamefont {S.}~\bibnamefont {Diehl}},\
  and\ \bibinfo {author} {\bibfnamefont {M.}~\bibnamefont {Buchhold}},\ }\href
  {https://doi.org/10.1103/PhysRevResearch.4.033001} {\bibfield  {journal}
  {\bibinfo  {journal} {Phys. Rev. Research}\ }\textbf {\bibinfo {volume}
  {4}},\ \bibinfo {pages} {033001} (\bibinfo {year} {2022})}\BibitemShut
  {NoStop}%
\bibitem [{\citenamefont {Ippoliti}\ \emph {et~al.}(2021)\citenamefont
  {Ippoliti}, \citenamefont {Gullans}, \citenamefont {Gopalakrishnan},
  \citenamefont {Huse},\ and\ \citenamefont {Khemani}}]{Ippoliti2021}%
  \BibitemOpen
  \bibfield  {author} {\bibinfo {author} {\bibfnamefont {M.}~\bibnamefont
  {Ippoliti}}, \bibinfo {author} {\bibfnamefont {M.~J.}\ \bibnamefont
  {Gullans}}, \bibinfo {author} {\bibfnamefont {S.}~\bibnamefont
  {Gopalakrishnan}}, \bibinfo {author} {\bibfnamefont {D.~A.}\ \bibnamefont
  {Huse}},\ and\ \bibinfo {author} {\bibfnamefont {V.}~\bibnamefont
  {Khemani}},\ }\href {https://doi.org/10.1103/PhysRevX.11.011030} {\bibfield
  {journal} {\bibinfo  {journal} {Phys. Rev. X}\ }\textbf {\bibinfo {volume}
  {11}},\ \bibinfo {pages} {011030} (\bibinfo {year} {2021})}\BibitemShut
  {NoStop}%
\bibitem [{\citenamefont {Lang}\ and\ \citenamefont
  {B{\ifmmode\ddot{u}\else\"{u}\fi}chler}(2020)}]{Lang2020Sep}%
  \BibitemOpen
  \bibfield  {author} {\bibinfo {author} {\bibfnamefont {N.}~\bibnamefont
  {Lang}}\ and\ \bibinfo {author} {\bibfnamefont {H.~P.}\ \bibnamefont
  {B{\ifmmode\ddot{u}\else\"{u}\fi}chler}},\ }\href
  {https://doi.org/10.1103/PhysRevB.102.094204} {\bibfield  {journal} {\bibinfo
   {journal} {Phys. Rev. B}\ }\textbf {\bibinfo {volume} {102}},\ \bibinfo
  {pages} {094204} (\bibinfo {year} {2020})}\BibitemShut {NoStop}%
\bibitem [{\citenamefont {{Lavasani}}\ \emph {et~al.}(2022)\citenamefont
  {{Lavasani}}, \citenamefont {{Luo}},\ and\ \citenamefont
  {{Vijay}}}]{Lavasani2022}%
  \BibitemOpen
  \bibfield  {author} {\bibinfo {author} {\bibfnamefont {A.}~\bibnamefont
  {{Lavasani}}}, \bibinfo {author} {\bibfnamefont {Z.-X.}\ \bibnamefont
  {{Luo}}},\ and\ \bibinfo {author} {\bibfnamefont {S.}~\bibnamefont
  {{Vijay}}},\ }\href@noop {} {\bibfield  {journal} {\bibinfo  {journal} {arXiv
  e-prints}\ } (\bibinfo {year} {2022})},\ \Eprint
  {https://arxiv.org/abs/2207.02877} {arXiv:2207.02877} \BibitemShut {NoStop}%
\bibitem [{\citenamefont {{Zhu}}\ \emph {et~al.}(2023)\citenamefont {{Zhu}},
  \citenamefont {{Tantivasadakarn}},\ and\ \citenamefont
  {{Trebst}}}]{Guoyi2023}%
  \BibitemOpen
  \bibfield  {author} {\bibinfo {author} {\bibfnamefont {G.-Y.}\ \bibnamefont
  {{Zhu}}}, \bibinfo {author} {\bibfnamefont {N.}~\bibnamefont
  {{Tantivasadakarn}}},\ and\ \bibinfo {author} {\bibfnamefont
  {S.}~\bibnamefont {{Trebst}}},\ }\href
  {https://doi.org/10.48550/arXiv.2303.17627} {\bibfield  {journal} {\bibinfo
  {journal} {arXiv e-prints}\ ,\ \bibinfo {eid} {arXiv:2303.17627}} (\bibinfo
  {year} {2023})}\BibitemShut {NoStop}%
\bibitem [{\citenamefont {Sriram}\ \emph {et~al.}(2023)\citenamefont {Sriram},
  \citenamefont {Rakovszky}, \citenamefont {Khemani},\ and\ \citenamefont
  {Ippoliti}}]{Sriram2022}%
  \BibitemOpen
  \bibfield  {author} {\bibinfo {author} {\bibfnamefont {A.}~\bibnamefont
  {Sriram}}, \bibinfo {author} {\bibfnamefont {T.}~\bibnamefont {Rakovszky}},
  \bibinfo {author} {\bibfnamefont {V.}~\bibnamefont {Khemani}},\ and\ \bibinfo
  {author} {\bibfnamefont {M.}~\bibnamefont {Ippoliti}},\ }\href
  {https://doi.org/10.1103/PhysRevB.108.094304} {\bibfield  {journal} {\bibinfo
   {journal} {Phys. Rev. B}\ }\textbf {\bibinfo {volume} {108}},\ \bibinfo
  {pages} {094304} (\bibinfo {year} {2023})}\BibitemShut {NoStop}%
\bibitem [{\citenamefont {{Li}}\ and\ \citenamefont
  {{Fisher}}(2021)}]{LiDecoding}%
  \BibitemOpen
  \bibfield  {author} {\bibinfo {author} {\bibfnamefont {Y.}~\bibnamefont
  {{Li}}}\ and\ \bibinfo {author} {\bibfnamefont {M.~P.~A.}\ \bibnamefont
  {{Fisher}}},\ }\href@noop {} {\bibfield  {journal} {\bibinfo  {journal}
  {arXiv e-prints}\ } (\bibinfo {year} {2021})},\ \Eprint
  {https://arxiv.org/abs/2108.04274} {arXiv:2108.04274} \BibitemShut {NoStop}%
\bibitem [{\citenamefont {Klocke}\ and\ \citenamefont
  {Buchhold}(2022)}]{Klocke_2022}%
  \BibitemOpen
  \bibfield  {author} {\bibinfo {author} {\bibfnamefont {K.}~\bibnamefont
  {Klocke}}\ and\ \bibinfo {author} {\bibfnamefont {M.}~\bibnamefont
  {Buchhold}},\ }\href {https://doi.org/10.1103/PhysRevB.106.104307} {\bibfield
   {journal} {\bibinfo  {journal} {Phys. Rev. B}\ }\textbf {\bibinfo {volume}
  {106}},\ \bibinfo {pages} {104307} (\bibinfo {year} {2022})}\BibitemShut
  {NoStop}%
\bibitem [{\citenamefont {Klocke}\ and\ \citenamefont
  {Buchhold}(2023)}]{Klocke2023}%
  \BibitemOpen
  \bibfield  {author} {\bibinfo {author} {\bibfnamefont {K.}~\bibnamefont
  {Klocke}}\ and\ \bibinfo {author} {\bibfnamefont {M.}~\bibnamefont
  {Buchhold}},\ }\href {https://doi.org/10.1103/PhysRevX.13.041028} {\bibfield
  {journal} {\bibinfo  {journal} {Phys. Rev. X}\ }\textbf {\bibinfo {volume}
  {13}},\ \bibinfo {pages} {041028} (\bibinfo {year} {2023})}\BibitemShut
  {NoStop}%
\bibitem [{\citenamefont {{Negari}}\ \emph {et~al.}(2023)\citenamefont
  {{Negari}}, \citenamefont {{Sahu}},\ and\ \citenamefont
  {{Hsieh}}}]{Negari2023}%
  \BibitemOpen
  \bibfield  {author} {\bibinfo {author} {\bibfnamefont {A.-R.}\ \bibnamefont
  {{Negari}}}, \bibinfo {author} {\bibfnamefont {S.}~\bibnamefont {{Sahu}}},\
  and\ \bibinfo {author} {\bibfnamefont {T.~H.}\ \bibnamefont {{Hsieh}}},\
  }\href {https://doi.org/10.48550/arXiv.2307.02292} {\bibfield  {journal}
  {\bibinfo  {journal} {arXiv e-prints}\ ,\ \bibinfo {eid} {arXiv:2307.02292}}
  (\bibinfo {year} {2023})}\BibitemShut {NoStop}%
\bibitem [{\citenamefont {Fan}\ \emph {et~al.}(2021)\citenamefont {Fan},
  \citenamefont {Vijay}, \citenamefont {Vishwanath},\ and\ \citenamefont
  {You}}]{Fan2021May}%
  \BibitemOpen
  \bibfield  {author} {\bibinfo {author} {\bibfnamefont {R.}~\bibnamefont
  {Fan}}, \bibinfo {author} {\bibfnamefont {S.}~\bibnamefont {Vijay}}, \bibinfo
  {author} {\bibfnamefont {A.}~\bibnamefont {Vishwanath}},\ and\ \bibinfo
  {author} {\bibfnamefont {Y.-Z.}\ \bibnamefont {You}},\ }\href
  {https://doi.org/10.1103/PhysRevB.103.174309} {\bibfield  {journal} {\bibinfo
   {journal} {Phys. Rev. B}\ }\textbf {\bibinfo {volume} {103}},\ \bibinfo
  {pages} {174309} (\bibinfo {year} {2021})}\BibitemShut {NoStop}%
\bibitem [{\citenamefont {Li}\ and\ \citenamefont {Fisher}(2021)}]{Li2021Mar}%
  \BibitemOpen
  \bibfield  {author} {\bibinfo {author} {\bibfnamefont {Y.}~\bibnamefont
  {Li}}\ and\ \bibinfo {author} {\bibfnamefont {M.~P.~A.}\ \bibnamefont
  {Fisher}},\ }\href {https://doi.org/10.1103/PhysRevB.103.104306} {\bibfield
  {journal} {\bibinfo  {journal} {Phys. Rev. B}\ }\textbf {\bibinfo {volume}
  {103}},\ \bibinfo {pages} {104306} (\bibinfo {year} {2021})}\BibitemShut
  {NoStop}%
\bibitem [{\citenamefont {{Behrends}}\ \emph {et~al.}(2022)\citenamefont
  {{Behrends}}, \citenamefont {{Venn}},\ and\ \citenamefont
  {{B{\'e}ri}}}]{Behrends2022}%
  \BibitemOpen
  \bibfield  {author} {\bibinfo {author} {\bibfnamefont {J.}~\bibnamefont
  {{Behrends}}}, \bibinfo {author} {\bibfnamefont {F.}~\bibnamefont {{Venn}}},\
  and\ \bibinfo {author} {\bibfnamefont {B.}~\bibnamefont {{B{\'e}ri}}},\
  }\href {https://doi.org/10.48550/arXiv.2212.08084} {\bibfield  {journal}
  {\bibinfo  {journal} {arXiv e-prints}\ ,\ \bibinfo {eid} {arXiv:2212.08084}}
  (\bibinfo {year} {2022})}\BibitemShut {NoStop}%
\bibitem [{\citenamefont {Hashizume}\ \emph {et~al.}(2022)\citenamefont
  {Hashizume}, \citenamefont {Bentsen},\ and\ \citenamefont
  {Daley}}]{Hashizume2022Mar}%
  \BibitemOpen
  \bibfield  {author} {\bibinfo {author} {\bibfnamefont {T.}~\bibnamefont
  {Hashizume}}, \bibinfo {author} {\bibfnamefont {G.}~\bibnamefont {Bentsen}},\
  and\ \bibinfo {author} {\bibfnamefont {A.~J.}\ \bibnamefont {Daley}},\ }\href
  {https://doi.org/10.1103/PhysRevResearch.4.013174} {\bibfield  {journal}
  {\bibinfo  {journal} {Phys. Rev. Res.}\ }\textbf {\bibinfo {volume} {4}},\
  \bibinfo {pages} {013174} (\bibinfo {year} {2022})}\BibitemShut {NoStop}%
\bibitem [{\citenamefont {Lang}\ and\ \citenamefont
  {B\"uchler}(2015)}]{Lang2015Jul}%
  \BibitemOpen
  \bibfield  {author} {\bibinfo {author} {\bibfnamefont {N.}~\bibnamefont
  {Lang}}\ and\ \bibinfo {author} {\bibfnamefont {H.~P.}\ \bibnamefont
  {B\"uchler}},\ }\href {https://doi.org/10.1103/PhysRevA.92.012128} {\bibfield
   {journal} {\bibinfo  {journal} {Phys. Rev. A}\ }\textbf {\bibinfo {volume}
  {92}},\ \bibinfo {pages} {012128} (\bibinfo {year} {2015})}\BibitemShut
  {NoStop}%
\bibitem [{\citenamefont {Lavasani}\ \emph
  {et~al.}(2021{\natexlab{b}})\citenamefont {Lavasani}, \citenamefont
  {Alavirad},\ and\ \citenamefont {Barkeshli}}]{Lavasani2021Dec}%
  \BibitemOpen
  \bibfield  {author} {\bibinfo {author} {\bibfnamefont {A.}~\bibnamefont
  {Lavasani}}, \bibinfo {author} {\bibfnamefont {Y.}~\bibnamefont {Alavirad}},\
  and\ \bibinfo {author} {\bibfnamefont {M.}~\bibnamefont {Barkeshli}},\ }\href
  {https://doi.org/10.1103/PhysRevLett.127.235701} {\bibfield  {journal}
  {\bibinfo  {journal} {Phys. Rev. Lett.}\ }\textbf {\bibinfo {volume} {127}},\
  \bibinfo {pages} {235701} (\bibinfo {year} {2021}{\natexlab{b}})}\BibitemShut
  {NoStop}%
\bibitem [{\citenamefont {Fowler}\ \emph {et~al.}(2012)\citenamefont {Fowler},
  \citenamefont {Mariantoni}, \citenamefont {Martinis},\ and\ \citenamefont
  {Cleland}}]{PhysRevA.86.032324}%
  \BibitemOpen
  \bibfield  {author} {\bibinfo {author} {\bibfnamefont {A.~G.}\ \bibnamefont
  {Fowler}}, \bibinfo {author} {\bibfnamefont {M.}~\bibnamefont {Mariantoni}},
  \bibinfo {author} {\bibfnamefont {J.~M.}\ \bibnamefont {Martinis}},\ and\
  \bibinfo {author} {\bibfnamefont {A.~N.}\ \bibnamefont {Cleland}},\ }\href
  {https://doi.org/10.1103/PhysRevA.86.032324} {\bibfield  {journal} {\bibinfo
  {journal} {Phys. Rev. A}\ }\textbf {\bibinfo {volume} {86}},\ \bibinfo
  {pages} {032324} (\bibinfo {year} {2012})}\BibitemShut {NoStop}%
\bibitem [{\citenamefont {Grassl}\ \emph {et~al.}(1997)\citenamefont {Grassl},
  \citenamefont {Beth},\ and\ \citenamefont {Pellizzari}}]{Grassl1997}%
  \BibitemOpen
  \bibfield  {author} {\bibinfo {author} {\bibfnamefont {M.}~\bibnamefont
  {Grassl}}, \bibinfo {author} {\bibfnamefont {T.}~\bibnamefont {Beth}},\ and\
  \bibinfo {author} {\bibfnamefont {T.}~\bibnamefont {Pellizzari}},\ }\href
  {https://doi.org/10.1103/PhysRevA.56.33} {\bibfield  {journal} {\bibinfo
  {journal} {Phys. Rev. A}\ }\textbf {\bibinfo {volume} {56}},\ \bibinfo
  {pages} {33} (\bibinfo {year} {1997})}\BibitemShut {NoStop}%
\bibitem [{\citenamefont {Stricker}\ \emph {et~al.}(2020)\citenamefont
  {Stricker}, \citenamefont {Vodola}, \citenamefont {Erhard}, \citenamefont
  {Postler}, \citenamefont {Meth}, \citenamefont {Ringbauer}, \citenamefont
  {Schindler}, \citenamefont {Monz}, \citenamefont
  {M{\ifmmode\ddot{u}\else\"{u}\fi}ller},\ and\ \citenamefont
  {Blatt}}]{Stricker2020Sep}%
  \BibitemOpen
  \bibfield  {author} {\bibinfo {author} {\bibfnamefont {R.}~\bibnamefont
  {Stricker}}, \bibinfo {author} {\bibfnamefont {D.}~\bibnamefont {Vodola}},
  \bibinfo {author} {\bibfnamefont {A.}~\bibnamefont {Erhard}}, \bibinfo
  {author} {\bibfnamefont {L.}~\bibnamefont {Postler}}, \bibinfo {author}
  {\bibfnamefont {M.}~\bibnamefont {Meth}}, \bibinfo {author} {\bibfnamefont
  {M.}~\bibnamefont {Ringbauer}}, \bibinfo {author} {\bibfnamefont
  {P.}~\bibnamefont {Schindler}}, \bibinfo {author} {\bibfnamefont
  {T.}~\bibnamefont {Monz}}, \bibinfo {author} {\bibfnamefont {M.}~\bibnamefont
  {M{\ifmmode\ddot{u}\else\"{u}\fi}ller}},\ and\ \bibinfo {author}
  {\bibfnamefont {R.}~\bibnamefont {Blatt}},\ }\href
  {https://doi.org/10.1038/s41586-020-2667-0} {\bibfield  {journal} {\bibinfo
  {journal} {Nature}\ }\textbf {\bibinfo {volume} {585}},\ \bibinfo {pages}
  {207} (\bibinfo {year} {2020})}\BibitemShut {NoStop}%
\bibitem [{\citenamefont {Krinner}\ \emph {et~al.}(2022)\citenamefont
  {Krinner}, \citenamefont {Lacroix}, \citenamefont {Remm}, \citenamefont
  {Di~Paolo}, \citenamefont {Genois}, \citenamefont {Leroux}, \citenamefont
  {Hellings}, \citenamefont {Lazar}, \citenamefont {Swiadek}, \citenamefont
  {Herrmann}, \citenamefont {Norris}, \citenamefont {Andersen}, \citenamefont
  {M{\"u}ller}, \citenamefont {Blais}, \citenamefont {Eichler},\ and\
  \citenamefont {Wallraff}}]{krinner2022realizing}%
  \BibitemOpen
  \bibfield  {author} {\bibinfo {author} {\bibfnamefont {S.}~\bibnamefont
  {Krinner}}, \bibinfo {author} {\bibfnamefont {N.}~\bibnamefont {Lacroix}},
  \bibinfo {author} {\bibfnamefont {A.}~\bibnamefont {Remm}}, \bibinfo {author}
  {\bibfnamefont {A.}~\bibnamefont {Di~Paolo}}, \bibinfo {author}
  {\bibfnamefont {E.}~\bibnamefont {Genois}}, \bibinfo {author} {\bibfnamefont
  {C.}~\bibnamefont {Leroux}}, \bibinfo {author} {\bibfnamefont
  {C.}~\bibnamefont {Hellings}}, \bibinfo {author} {\bibfnamefont
  {S.}~\bibnamefont {Lazar}}, \bibinfo {author} {\bibfnamefont
  {F.}~\bibnamefont {Swiadek}}, \bibinfo {author} {\bibfnamefont
  {J.}~\bibnamefont {Herrmann}}, \bibinfo {author} {\bibfnamefont {G.~J.}\
  \bibnamefont {Norris}}, \bibinfo {author} {\bibfnamefont {C.~K.}\
  \bibnamefont {Andersen}}, \bibinfo {author} {\bibfnamefont {M.}~\bibnamefont
  {M{\"u}ller}}, \bibinfo {author} {\bibfnamefont {A.}~\bibnamefont {Blais}},
  \bibinfo {author} {\bibfnamefont {C.}~\bibnamefont {Eichler}},\ and\ \bibinfo
  {author} {\bibfnamefont {A.}~\bibnamefont {Wallraff}},\ }\href
  {https://doi.org/https://doi.org/10.1038/s41586-022-04566-8} {\bibfield
  {journal} {\bibinfo  {journal} {Nature}\ }\textbf {\bibinfo {volume} {605}},\
  \bibinfo {pages} {669} (\bibinfo {year} {2022})}\BibitemShut {NoStop}%
\bibitem [{\citenamefont {Zhao}\ \emph {et~al.}(2022)\citenamefont {Zhao},
  \citenamefont {Ye}, \citenamefont {Huang}, \citenamefont {Zhang},
  \citenamefont {Wu}, \citenamefont {Guan}, \citenamefont {Zhu}, \citenamefont
  {Wei}, \citenamefont {He}, \citenamefont {Cao}, \citenamefont {Chen},
  \citenamefont {Chung}, \citenamefont {Deng}, \citenamefont {Fan},
  \citenamefont {Gong}, \citenamefont {Guo}, \citenamefont {Guo}, \citenamefont
  {Han}, \citenamefont {Li}, \citenamefont {Li}, \citenamefont {Li},
  \citenamefont {Liang}, \citenamefont {Lin}, \citenamefont {Qian},
  \citenamefont {Rong}, \citenamefont {Su}, \citenamefont {Sun}, \citenamefont
  {Wang}, \citenamefont {Wu}, \citenamefont {Xu}, \citenamefont {Ying},
  \citenamefont {Yu}, \citenamefont {Zha}, \citenamefont {Zhang}, \citenamefont
  {Huo}, \citenamefont {Lu}, \citenamefont {Peng}, \citenamefont {Zhu},\ and\
  \citenamefont {Pan}}]{zhao2022realization}%
  \BibitemOpen
  \bibfield  {author} {\bibinfo {author} {\bibfnamefont {Y.}~\bibnamefont
  {Zhao}}, \bibinfo {author} {\bibfnamefont {Y.}~\bibnamefont {Ye}}, \bibinfo
  {author} {\bibfnamefont {H.-L.}\ \bibnamefont {Huang}}, \bibinfo {author}
  {\bibfnamefont {Y.}~\bibnamefont {Zhang}}, \bibinfo {author} {\bibfnamefont
  {D.}~\bibnamefont {Wu}}, \bibinfo {author} {\bibfnamefont {H.}~\bibnamefont
  {Guan}}, \bibinfo {author} {\bibfnamefont {Q.}~\bibnamefont {Zhu}}, \bibinfo
  {author} {\bibfnamefont {Z.}~\bibnamefont {Wei}}, \bibinfo {author}
  {\bibfnamefont {T.}~\bibnamefont {He}}, \bibinfo {author} {\bibfnamefont
  {S.}~\bibnamefont {Cao}}, \bibinfo {author} {\bibfnamefont {F.}~\bibnamefont
  {Chen}}, \bibinfo {author} {\bibfnamefont {T.-H.}\ \bibnamefont {Chung}},
  \bibinfo {author} {\bibfnamefont {H.}~\bibnamefont {Deng}}, \bibinfo {author}
  {\bibfnamefont {D.}~\bibnamefont {Fan}}, \bibinfo {author} {\bibfnamefont
  {M.}~\bibnamefont {Gong}}, \bibinfo {author} {\bibfnamefont {C.}~\bibnamefont
  {Guo}}, \bibinfo {author} {\bibfnamefont {S.}~\bibnamefont {Guo}}, \bibinfo
  {author} {\bibfnamefont {L.}~\bibnamefont {Han}}, \bibinfo {author}
  {\bibfnamefont {N.}~\bibnamefont {Li}}, \bibinfo {author} {\bibfnamefont
  {S.}~\bibnamefont {Li}}, \bibinfo {author} {\bibfnamefont {Y.}~\bibnamefont
  {Li}}, \bibinfo {author} {\bibfnamefont {F.}~\bibnamefont {Liang}}, \bibinfo
  {author} {\bibfnamefont {J.}~\bibnamefont {Lin}}, \bibinfo {author}
  {\bibfnamefont {H.}~\bibnamefont {Qian}}, \bibinfo {author} {\bibfnamefont
  {H.}~\bibnamefont {Rong}}, \bibinfo {author} {\bibfnamefont {H.}~\bibnamefont
  {Su}}, \bibinfo {author} {\bibfnamefont {L.}~\bibnamefont {Sun}}, \bibinfo
  {author} {\bibfnamefont {S.}~\bibnamefont {Wang}}, \bibinfo {author}
  {\bibfnamefont {Y.}~\bibnamefont {Wu}}, \bibinfo {author} {\bibfnamefont
  {Y.}~\bibnamefont {Xu}}, \bibinfo {author} {\bibfnamefont {C.}~\bibnamefont
  {Ying}}, \bibinfo {author} {\bibfnamefont {J.}~\bibnamefont {Yu}}, \bibinfo
  {author} {\bibfnamefont {C.}~\bibnamefont {Zha}}, \bibinfo {author}
  {\bibfnamefont {K.}~\bibnamefont {Zhang}}, \bibinfo {author} {\bibfnamefont
  {Y.-H.}\ \bibnamefont {Huo}}, \bibinfo {author} {\bibfnamefont {C.-Y.}\
  \bibnamefont {Lu}}, \bibinfo {author} {\bibfnamefont {C.-Z.}\ \bibnamefont
  {Peng}}, \bibinfo {author} {\bibfnamefont {X.}~\bibnamefont {Zhu}},\ and\
  \bibinfo {author} {\bibfnamefont {J.-W.}\ \bibnamefont {Pan}},\ }\href
  {https://doi.org/10.1103/PhysRevLett.129.030501} {\bibfield  {journal}
  {\bibinfo  {journal} {Phys. Rev. Lett.}\ }\textbf {\bibinfo {volume} {129}},\
  \bibinfo {pages} {030501} (\bibinfo {year} {2022})}\BibitemShut {NoStop}%
\bibitem [{\citenamefont {{Google Quantum AI}}(2023)}]{google2023suppressing}%
  \BibitemOpen
  \bibfield  {author} {\bibinfo {author} {\bibnamefont {{Google Quantum AI}}},\
  }\href {https://doi.org/https://doi.org/10.1038/s41586-022-05434-1}
  {\bibfield  {journal} {\bibinfo  {journal} {Nature}\ }\textbf {\bibinfo
  {volume} {614}},\ \bibinfo {pages} {676} (\bibinfo {year}
  {2023})}\BibitemShut {NoStop}%
\bibitem [{\citenamefont {Dennis}\ \emph {et~al.}(2002)\citenamefont {Dennis},
  \citenamefont {Kitaev}, \citenamefont {Landahl},\ and\ \citenamefont
  {Preskill}}]{Dennis2002Sep}%
  \BibitemOpen
  \bibfield  {author} {\bibinfo {author} {\bibfnamefont {E.}~\bibnamefont
  {Dennis}}, \bibinfo {author} {\bibfnamefont {A.}~\bibnamefont {Kitaev}},
  \bibinfo {author} {\bibfnamefont {A.}~\bibnamefont {Landahl}},\ and\ \bibinfo
  {author} {\bibfnamefont {J.}~\bibnamefont {Preskill}},\ }\href
  {https://doi.org/10.1063/1.1499754} {\bibfield  {journal} {\bibinfo
  {journal} {J. Math. Phys.}\ }\textbf {\bibinfo {volume} {43}},\ \bibinfo
  {pages} {4452} (\bibinfo {year} {2002})}\BibitemShut {NoStop}%
\bibitem [{Note1()}]{Note1}%
  \BibitemOpen
  \bibinfo {note} {Without loss of generality we assume the measurement outcome
  $+1$.}\BibitemShut {Stop}%
\bibitem [{\citenamefont {Aaronson}\ and\ \citenamefont
  {Gottesman}(2004)}]{Aaronson_2004}%
  \BibitemOpen
  \bibfield  {author} {\bibinfo {author} {\bibfnamefont {S.}~\bibnamefont
  {Aaronson}}\ and\ \bibinfo {author} {\bibfnamefont {D.}~\bibnamefont
  {Gottesman}},\ }\href {http://dx.doi.org/10.1103/PhysRevA.70.052328}
  {\bibfield  {journal} {\bibinfo  {journal} {Phys. Rev. A}\ }\textbf {\bibinfo
  {volume} {70}} (\bibinfo {year} {2004})}\BibitemShut {NoStop}%
\bibitem [{\citenamefont {Stauffer}\ and\ \citenamefont
  {Aharony}(2017)}]{DietrichStauffer2017Jan}%
  \BibitemOpen
  \bibfield  {author} {\bibinfo {author} {\bibfnamefont {D.}~\bibnamefont
  {Stauffer}}\ and\ \bibinfo {author} {\bibfnamefont {A.}~\bibnamefont
  {Aharony}},\ }\href {https://doi.org/10.1201/9781315274386} {\emph {\bibinfo
  {title} {{Introduction To Percolation Theory:Second Edition}}}}\ (\bibinfo
  {publisher} {Taylor {\&} Francis},\ \bibinfo {address} {Andover, England,
  UK},\ \bibinfo {year} {2017})\BibitemShut {NoStop}%
\bibitem [{\citenamefont {Stace}\ \emph {et~al.}(2009)\citenamefont {Stace},
  \citenamefont {Barrett},\ and\ \citenamefont {Doherty}}]{Stace2009May}%
  \BibitemOpen
  \bibfield  {author} {\bibinfo {author} {\bibfnamefont {T.~M.}\ \bibnamefont
  {Stace}}, \bibinfo {author} {\bibfnamefont {S.~D.}\ \bibnamefont {Barrett}},\
  and\ \bibinfo {author} {\bibfnamefont {A.~C.}\ \bibnamefont {Doherty}},\
  }\href {https://doi.org/10.1103/PhysRevLett.102.200501} {\bibfield  {journal}
  {\bibinfo  {journal} {Phys. Rev. Lett.}\ }\textbf {\bibinfo {volume} {102}},\
  \bibinfo {pages} {200501} (\bibinfo {year} {2009})}\BibitemShut {NoStop}%
\bibitem [{Note2()}]{Note2}%
  \BibitemOpen
  \bibinfo {note} {We note that a similar phase diagram has been obtained in
  the measurement-based quantum computation setup proposed in Ref.~\cite
  {Negari2023}: In this study, Pauli-measurements evolve the bulk of a 2D toric
  code ground state in order to generate entanglement at its one-dimensional
  spatial boundary. While the $\protect \hat X$ or $\protect \hat Z$ dominated
  regimes yield an area law entangled boundary, the $\protect \hat
  Y$-measurement dominated regime induces a $\log ^2 (L)$-growth of
  entanglement at a boundary of length $L$.}\BibitemShut {Stop}%
\end{thebibliography}%

\end{document}